\documentclass[aps,prb,preprint, amsmath,amssymb, groupedaddress]{revtex4-2}

\usepackage{txfonts}
\usepackage{bm}
\usepackage{graphicx}
\usepackage{dcolumn}
\usepackage{color}

\begin{document}


\newcommand{\ogata}[1]{\textcolor{black}{#1}}

\title{Theory of Phason Drag Effect on Thermoelectricity} 


\author{Hidetoshi Fukuyama}
\affiliation{Tokyo University of Science, Shinjuku, Tokyo 162-8601}

\author{Masao Ogata}
\email[]{ogata@phys.s.u-tokyo.ac.jp}
\affiliation{Department of Physics, University of Tokyo,
Bunkyo, Tokyo 113-0033, Japan }


\date{\today}

\begin{abstract}
Lee, Rice and Anderson, in their monumental paper, have proved the existence of a collective mode 
describing the coupled motion of electron density and phonons in one-dimensional 
incommensurate charge density wave (CDW) in the Peierls state. 
This mode, which represents the coherent sliding motion of electrons and lattice distortions and 
affects low energy transport properties, is described by the phase of the complex order parameter of the 
Peierls condensate, leading to Fr\"ohlich superconductivity in pure systems.
Once spatial disorder is present, however, phason is pinned and system is transformed into an insulating ground state: 
a dramatic change. 
Since phason can be considered as an ultimate of phonon drag effect, it is of interest to see its effects
on thermoelectricity, which has been studied in the present paper based linear response theory of Kubo and Luttinger.
The result indicates that a large absolute value of Seebeck coefficient proportional 
to the square root of resistivity is expected at low temperatures $k_{\rm B}T/\Delta <<1$ ($\Delta$: Peierls gap) 
with opposite sign to the electronic contributions in the absence of Peierls gap. 
\end{abstract}


\maketitle

\section{Introduction}

Various aspects of thermoelectric effect have been extensively studied so far 
both theoretically and experimentally.\cite{ExpRev,ExpRev2}
Especially recent social needs reflecting the fact that quite a fraction of primary energy 
is wasted as heat strongly urge quest of materials with high thermoelectric capability. 
Motivated by this understanding, we have been developing studies toward systematic 
understanding of thermoelectricity beyond Boltzmann transport theory 
based on linear response theory of Kubo\cite{Kubo} and Luttinger.\cite{Luttinger} 
Those include spin-Seebeck effect free from contamination of electric current,\cite{OF1} 
identification of the range of validity of Sommerfeld-Bethe (SB) relation,\cite{OFSB}
phonon drag effect in the presence of impurity band,\cite{Matsuura}
and n-type and bipolar carbon-nanotubes indicating importance of band-edge engineering and
possible probing of morphology of samples in experiments.\cite{YamaFuku1,YamaFuku2} 
In this paper phason drag effects are studied as an example of phonon drag, 
which has long been known in doped semiconductors to play important roles\cite{Gurevich,Herring,Mahan,Zhou} 
and was proposed very recently to be the case also in FeSb$_2$.\cite{Matsuura}

Phason is the collective mode of the electron-phonon coupled systems in the incommensurate Peierls phase 
resulting in charge density wave (CDW), where spatially-modulated electron density and 
lattice distortion are locked with the same periodicity. 
Lee, Rice and Anderson (LRA)\cite{LRA} discovered this phason in view of experimental finding of extraordinary 
conductivity in TTF-TCNQ leading to the controversial discussions of possible 
Fr\"ohlich superconductivity.\cite{Coleman} 
When CDW moves, electrons and lattice distortion move together (sliding mode) and the dynamics is 
described by the phase of complex order parameter of Peierls phase, then called as phason. 
Hence phason is considered to be the ultimate form of phonon drag. 
In contrast to the case of superconductivity, phason, which is due to 
diagonal long range order, is sensitive to spatial inhomogeneity resulting in impurity pinning. 
Once phason is pinned, there is no sliding and the state is insulating at absolute zero. 
However at finite temperature, pinned phason will move locally by the creation of soliton pairs 
induced by the thermal excitation leading to the activation-type temperature dependence of 
conductivity, $L_{11}$.\cite{Cohen}
In this paper we study thermoelectric conductivity, $L_{12}$, and the Seebeck coefficient 
$S=L_{12}/TL_{11}$, with $T$ being temperature, due to phason drag at such low temperature region 
for one-dimensional electron-phonon Peierls phase by use of thermal Green function. 

Regarding phason contributions to $L_{12}$, Yoshimoto and Kurihara\cite{Kurihara} 
studied electronic contribution in clean systems without disorder. 
In this paper, we explore the phason drag contributions in the presence of impurity pinning. 

In Section 2, we introduce one-dimensional electron-phonon system and phason. 
The formulation by Lee, Rice, and Anderson\cite{LRA} is modified in accordance with the present framework. 
In Section 3, the electrical conductivity due to phason is discussed, and 
in Section 4, results of the phason drag contribution to $L_{12}$ and the resulting Seebeck 
coefficient are given.  Section 5 is devoted to summary.

\section{One-dimensional electron-phonon system and phason}

\ogata{
We consider one-dimensional electron-phonon system by Fr\"ohlich model, $H_0$, to describe Peierls transition
in the presence of random distribution of impurities, $H'$, $H=H_0+H'$, where $H_0$ and $H'$ 
are given as follows:} 
\begin{eqnarray}
H_0&&= \sum_{p,\sigma} \varepsilon_p c^\dagger_{p,\sigma} c^{\phantom{\dagger}}_{p,\sigma} 
+ \sum_q \hbar \omega_q b^\dagger_q b^{\phantom{\dagger}}_q
+ \frac{1}{\sqrt{L}} \sum_{p,q,\sigma} g_q c^\dagger_{p+q, \sigma} c^{\phantom{\dagger}}_{p,\sigma} (b_q +b_{-q}^\dagger), \cr
\ogata{H'}&& \ogata{= \sum_i \int v(x-R_i) \rho(x) dx
= \frac{1}{L} \sum_i \sum_{p,q,\sigma} e^{-iq R_i} v_q c^\dagger_{p+q,\sigma} c^{\phantom{\dagger}}_{p,\sigma},}
\label{FrohlichHamil}
\end{eqnarray}
Here, $c^\dagger_{p,\sigma}$ and $b^\dagger_q$ are creation operators for a one-dimensional 
Bloch electron and a phonon with energies $\varepsilon_p$
and $\hbar \omega_q$, respectively, 
$L$ is the length of the system, $g_q$ represents the electron-phonon coupling constant, 
\ogata{
$R_i$ represents the position of impurities, 
$v(x)$ and $\rho(x)$ are impurity potential and electron density, respectively, and $v_q$ is the 
Fourier transform of $v(x)$.
First we focus on $H_0$ and  the effects of $H'$ will be treated later. }

In the mean-field theory of uniform Peierls phase, the lattice distortion is described by the order parameter
\begin{equation}
\Delta = \frac{1}{\sqrt{L}} g_Q \left( \langle b_Q \rangle + \langle b^\dagger_{-Q} \rangle \right)
\equiv e^{i\phi} \Delta_0,
\label{DeltaDef0}
\end{equation}
($\Delta_0>0$) with $Q=2k_{\rm F}$, 
and the mean-field Hamiltonian for electrons becomes
\begin{equation}
H_{\rm MF} = \sum_{k, \sigma}
\left( \begin{array}{cc} c^\dagger_{\frac{Q}{2}+k,\sigma}, & c^\dagger_{-\frac{Q}{2}+k, \sigma} \end{array} \right)
\left( \begin{array}{cc}
\xi_k & \Delta \\
\Delta^* & -\xi_k \\
\end{array} \right)
\left( \begin{array}{c} c_{\frac{Q}{2}+k,\sigma} \\ c_{-\frac{Q}{2}+k, \sigma} \\ \end{array} \right),
\label{MFHamilton}
\end{equation}
where $|k|<Q/2$, and we have linearized the energy dispersion as 
$\varepsilon_p -\mu \sim \xi_k$ in the vicinity of $k_{\rm F}=Q/2$ with $\xi_k = \hbar v_{\rm F} k$ and 
$k=p-k_{\rm F}$. 
Similarly in the vicinity of $-k_{\rm F}=-Q/2$, we have linearized as
$\varepsilon_p -\mu \sim -\xi_k$ with $k=p+k_{\rm F}$.
It is to be noted that we consider the cases where $v_{\rm F}$ is positive and negative. 
The self-consistency equation for $\Delta$ is 
\begin{equation}
\Delta = \frac{4g_Q^2}{\hbar \omega_Q L} \sum_k \frac{\Delta}{2E_k} \left\{ f(-E_k) -f(E_k) \right\},
\label{SCeqForD}
\end{equation}
where 
\begin{equation}
E_k = \sqrt{\xi_k^2 + |\Delta|^2},
\end{equation}
and $f(\varepsilon) = 1/(e^{\beta\varepsilon}+1)$ is the Fermi distribution function with $\beta=1/k_{\rm B}T$. 

As performed by LRA, $\Delta$ can be chosen as real (i.e., $\phi=0$) in the uniform mean-field solution 
by redefining the operator as 
$\tilde c^\dagger_{\frac{Q}{2}+k, \sigma}= c^\dagger_{\frac{Q}{2}+k, \sigma} e^{i\phi}$ while 
$c^\dagger_{-\frac{Q}{2}+k, \sigma}$ is not changed. 
However, we keep $\phi$ in the following since $\phi$ is no longer uniform in the presence of 
impurity pinning and the dynamics of phason is represented by the spatial and temporary dependence of $\phi$ 
as described in the phase Hamiltonian.\cite{Fukuyama76}
\ogata{In fact, the charge density is given as}
\begin{equation}
\ogata{\rho(x,t) = n_e + \rho_0 \cos(Qx+\phi(x,t)),} \label{RhoCDWdef}
\end{equation}
\ogata{where $n_e$ is the average electron density, $\rho_0=\hbar \omega_Q |\Delta|/2 g_Q^2$}
and $\partial \phi(x,t)/\partial t$ and $\partial \phi(x,t)/\partial x$ give
the electric current density and the local modulation of electric charge density, respectively.\cite{TakayamaFuku,FukuLee}

To study the phason mode and amplitude mode in the case of a constant $\phi$, 
we introduce phonon propagators in the matrix form
\begin{equation}
{\mathcal D}_{mn}(q,\tau) = -\langle T_\tau 
[\left( b^{\phantom{\dagger}}_{mQ+q} (\tau) + b^{{\dagger}}_{-mQ-q} (\tau) \right) 
\left( b^\dagger_{nQ+q} (0) + b^{\phantom{\dagger}}_{-nQ-q} (0)\right) ] \rangle, 
\label{PGreenDef}
\end{equation}
where $m, n=\pm$, and electron Green's functions
\begin{equation}
{\mathcal G}_{mn}(k,\tau) = -\langle T_\tau [c^{\phantom{\dagger}}_{m{Q}/{2}+k,\sigma} (\tau) c^\dagger_{n{Q}/{2}+k, \sigma} (0)] \rangle.
\end{equation}
For the mean-field Hamiltonian of eq.~(\ref{MFHamilton}), the Fourier transform of ${\mathcal G}_{mn}(k,\tau)$ is given by
\begin{equation}
{\mathcal G}(k, i\varepsilon_n) = \frac{1}{(i\varepsilon_n)^2-E_k^2} 
\left( \begin{array}{cc}
i\varepsilon_n + \xi_k & \Delta \cr \Delta^* & i\varepsilon_n - \xi_k 
\end{array} \right),
\label{GreenF}
\end{equation}
where $\varepsilon_n= (2n+1)\pi k_{\rm B}T$ is the Matsubara frequency ($n$ being an integer). 

As shown by LRA, the Dyson equation for $\mathcal D_{mn}$ leads to
\begin{equation}
{\mathcal D}_{++}(q, i\omega_\nu) \pm e^{-2i\phi} {\mathcal D}_{+-}(q, i\omega_\nu) = 
\frac{{\mathcal D}^{(0)}(i\omega_\nu) }{1-\left[\Pi_{++}(q, i\omega_\nu) \pm e^{-2i\phi} \Pi_{+-}(q, i\omega_\nu) \right] 
{\mathcal D}^{(0)}(i\omega_\nu)},
\label{Dyson}
\end{equation}
where $\omega_\nu = 2\pi \nu k_{\rm B}T$ is the Matsubara frequency ($\nu$ being an integer), and 
\begin{eqnarray}
\Pi_{++}(q, i\omega_\nu) &=& 2g_{Q}^2 \frac{k_{\rm B} T}{L} \sum_{k, n} 
{\mathcal G}_{++}(k+q, i\varepsilon_n+i\omega_\nu) {\mathcal G}_{--}(k, i\varepsilon_n), \cr
\Pi_{+-}(q, i\omega_\nu) &=& 2g_{Q}^2 \frac{k_{\rm B} T}{L} \sum_{k, n} 
{\mathcal G}_{+-}(k+q, i\varepsilon_n+i\omega_\nu) {\mathcal G}_{+-}(k, i\varepsilon_n),
\label{PiDef}
\end{eqnarray}
where $g_{Q+q}\sim g_Q$ has been assumed. 
(For completeness, the derivation of eq.~(\ref{Dyson}) is shown in Appendix A.)
It is to be noted that the zeroth order phonon propagator 
\begin{equation}
{\mathcal D}^{(0)}_{mn}(q,i\omega_\nu) = \delta_{mn}\frac{2\hbar \omega_{mQ+q}}{(i\omega_\nu)^2-\hbar^2 \omega_{mQ+q}^2} 
\end{equation}
has been approximated as
\begin{equation}
{\mathcal D}^{(0)}_{mn}(q,i\omega_\nu) \sim \delta_{mn} {\mathcal D}^{(0)}(i\omega_\nu) 
= \delta_{mn} \frac{2\hbar \omega_Q}{(i\omega_\nu)^2 - \hbar^2 \omega_Q^2}.
\label{D0Approxi}
\end{equation}

The denominator for ${\mathcal D}_{++}-e^{-2i\phi}{\mathcal D}_{+-}$ in eq.~(\ref{Dyson}) 
leads to $q$-linear mode, $\omega = v |q|$ ($v>0$), which is phason,
while that of ${\mathcal D}_{++}+e^{-2i\phi}{\mathcal D}_{+-}$ leads to amplitude mode. 
Therefore, phason and amplitude propagators are defined as
\begin{eqnarray}
P(q, i\omega_\nu) &=& {\mathcal D}_{++} (q, i\omega_\nu)- e^{-2i\phi} {\mathcal D}_{+-} (q, i\omega_\nu), \cr
A(q, i\omega_\nu) &=& {\mathcal D}_{++} (q, i\omega_\nu)+ e^{-2i\phi} {\mathcal D}_{+-} (q, i\omega_\nu),
\label{DefPhasonAmp}
\end{eqnarray}
respectively. 
For small-$q$ and small-$(i\omega_\nu)$ region, they become
\begin{eqnarray}
P(q, i\omega_\nu) &=& 
\frac{2\hbar\omega_Q/(1+X)}{(i\omega_\nu)^2 -(\hbar v q)^2}, \cr
A(q, i\omega_\nu) &=& \frac{2\hbar \omega_Q/(1+X/3)}{(i\omega_\nu)^2 - \hbar^2 \omega_{\rm am}^2(q)},
\label{DefPhasonPropXX}
\end{eqnarray}
where $X = \omega_Q g_Q^2 / 2\pi |v_{\rm F}| \Delta_0^2$, and the phason velocity $v$ is given by
\begin{equation}
v= (X/(1+X))^{1/2} |v_{\rm F}|, 
\label{PhasonVelocityXX}
\end{equation}
The dispersion of the amplitude mode, $\omega_{\rm am}(q)$, is 
\begin{equation}
\omega_{\rm am}(q) = \sqrt{\frac{4X}{1+\frac{X}{3}}\Delta_0^2 + \frac{X}{3+X}(\hbar v_{\rm F}q)^2}.
\end{equation}
The details of calculations are shown in Appendix A.

\section{Electrical conductivity due to phasons}

Kubo formula for dynamical electrical conductivity for uniform electric field, $\sigma(\omega)$ is given by
\begin{equation}
L_{11} = \frac{1}{i\omega} \left[ \Phi_{11}(i\omega_\lambda\rightarrow \hbar\omega+i\delta)-\Phi_{11}(0)\right],
\end{equation}
with
\begin{equation}
\Phi_{11}(i\omega_\lambda) = \frac{1}{L} \int_0^\beta d\tau \langle 
T_\tau [ J_e(\tau) J_e(0)] \rangle e^{i\omega_\lambda \tau}.
\label{Phi11Def}
\end{equation}
Here $J_e$ is the electronic current
\begin{equation}
J_e = -e \sum_{k, \sigma}
\left( \begin{array}{cc} c^\dagger_{\frac{Q}{2}+k,\sigma}, & c^\dagger_{-\frac{Q}{2}+k, \sigma} \end{array} \right)
v_{\rm F} \sigma_z
\left( \begin{array}{c} c_{\frac{Q}{2}+k,\sigma} \cr c_{-\frac{Q}{2}+k, \sigma} \end{array} \right),
\end{equation}
where $-e$ is the electron charge ($e>0$), $\sigma_z$ is the $z$-component of Pauli matrices. 
As shown by LRA, the conductivity due to phason is governed by processes in Fig.~\ref{fcondL11}.\cite{LRA}
For example, Fig.~\ref{fcondL11}(a) gives the following contribution to $\Phi_{11}(i\omega_\lambda)$:
\begin{eqnarray}
{\rm (a)}: \quad -g_Q^2 &&\frac{4e^2 v_{\rm F}^2 (k_{\rm B}T)^2}{L^2} \sum_{k,k', \varepsilon_n,\varepsilon'_n} 
{\rm Tr} \left[ \sigma_z {\mathcal G} (k, i\varepsilon_n+i\omega_\lambda) \sigma_+ {\mathcal G}(k, i\varepsilon_n) \right] \cr
&&\times {\mathcal D}_{++} (0, i\omega_\lambda) 
{\rm Tr} \left[ \sigma_- {\mathcal G} (k', i\varepsilon'_n+i\omega_\lambda) 
\sigma_z {\mathcal G}(k', i\varepsilon'_n) \right] \cr
\quad = - g_Q^2 &&\frac{4e^2 v_{\rm F}^2 (k_{\rm B}T)^2}{L^2} \sum_{k, \varepsilon_n} 
\ogata{
\left\{ {\mathcal G}_{++}(k, i\varepsilon_n+i\omega_\lambda) {\mathcal G}_{-+}(k, i\varepsilon_n) 
-{\mathcal G}_{-+}(k, i\varepsilon_n+i\omega_\lambda) {\mathcal G}_{--}(k, i\varepsilon_n) \right\} } \cr
&&\times {\mathcal D}_{++} (0, i\omega_\lambda) \sum_{k',\varepsilon'_n}
\ogata{
\left\{ {\mathcal G}_{++}(k', i\varepsilon_n'+i\omega_\lambda) {\mathcal G}_{+-}(k', i\varepsilon_n') 
-{\mathcal G}_{+-}(k', i\varepsilon_n'+i\omega_\lambda) {\mathcal G}_{--}(k', i\varepsilon_n') \right\} } \cr
\quad = - g_Q^2 &&\frac{4e^2 v_{\rm F}^2 (k_{\rm B}T)^2}{L^2} \sum_{k, \varepsilon_n}
\frac{ \Delta^* (i\omega_\lambda +2\xi_k )}
{\left\{ (i\varepsilon_n+i\omega_\lambda)^2-E_k^2 \right\} \left\{ (i\varepsilon_n)^2-E_k^2 \right\}} \cr
&&\times {\mathcal D}_{++} (0, i\omega_\lambda)  \sum_{k',\varepsilon'_n}  
\frac{ \Delta (i\omega_\lambda +2 \xi_{k'})}
{\left\{ (i\varepsilon'_n+i\omega_\lambda)^2-E_{k'}^2 \right\} \left\{ (i\varepsilon'_n)^2-E_{k'}^2 \right\}},
\end{eqnarray}
\ogata{where $\sigma_\pm = (\sigma_x \pm i \sigma_y)/2$ and $\sigma_x, \sigma_y, \sigma_z$ are $2\times 2$ Pauli matrices.}
In the last expression, the terms proportional to $\xi_k$ and $\xi_{k'}$ in the numerator vanish 
since they are odd functions of $k$ and $k'$, respectively. 
Figures \ref{fcondL11}(b)-(d) can be calculated similarly and their total becomes
\begin{eqnarray}
\Phi_{11}(i\omega_\lambda) &=& - {4e^2 v_{\rm F}^2 g_Q^2 \Delta_0^2 (i\omega_\lambda)^2} 
\left[ \frac{k_{\rm B}T}{L} \sum_{k, \varepsilon_n}
\frac{1}{\left\{ (i\varepsilon_n+i\omega_\lambda)^2-E_k^2 \right\} \left\{ (i\varepsilon_n)^2-E_k^2 \right\}} \right]^2\cr
&&\times \left\{ {\mathcal D}_{++} (0, i\omega_\lambda) - e^{-2i\phi} {\mathcal D}_{+-} (0, i\omega_\lambda) 
- e^{2i\phi} {\mathcal D}_{-+} (0, i\omega_\lambda) +{\mathcal D}_{--} (0, i\omega_\lambda) \right\}.
\label{CondTotal}
\end{eqnarray}
Noting that ${\mathcal D}_{--}={\mathcal D}_{++}$ and 
${\mathcal D}_{-+}=e^{-4i\phi} {\mathcal D}_{+-}$ (see Appendix A), we see that the last 
parentheses in eq.~(\ref{CondTotal}) is equal to the twice of the phason propagator $P(0, i\omega_\lambda)$. 

\begin{figure}[t]
\caption{Feynman diagrams for conductivity due to phason. 
The solid lines and the wavy lines represent electron and phonon Green's functions, respectively, 
The $\pm$ signs attached to the solid lines
represent the subscripts of electron Green's functions ${\mathcal G}_{mn}$ with $m, n=\pm$, 
and the $\pm Q$ attached to the wavy lines represent the subscripts of phonon Green's functions
${\mathcal D}_{mn}$.}
\label{fcondL11}
\end{figure}

In the lowest order of $i\omega_\lambda$ and $T\rightarrow 0$, the $k$-summation and 
the Matsubara frequency summation in eq.~(\ref{CondTotal}) can be carried out as
\begin{eqnarray}
\frac{k_{\rm B}T}{L} \sum_{k, \varepsilon_n} \frac{1}{\left\{ (i\varepsilon_n)^2-E_k^2 \right\}^2} 
&=& -\frac{1}{L} \sum_k \int \frac{dz}{2\pi i} f(z) \frac{1}{ \left( z^2-E_k^2 \right)^2} \cr
&=& \frac{1}{L} \sum_k \frac{1}{4E_k^3} 
=\frac{1}{4\pi \hbar |v_{\rm F}| \Delta_0^2}. 
\end{eqnarray}
Therefore the conductivity is given by 
\begin{equation}
\sigma(\omega)
= \frac{i\omega}{2} \left( \frac{e}{\pi} \right)^2 \frac{g_Q^2}{\Delta_0^2} P(0, \hbar\omega+i\delta). 
\label{phasonSigma}
\end{equation}
\ogata{Equation (\ref{phasonSigma}) together with eq.~(\ref{DefPhasonPropXX}) leads to
$\sigma(\omega)=n_e e^2/i\omega m^*$ with $m^*=((1+X)/X) m$ and $n_e=2k_F/\pi$, 
which is the result by Lee, Rice and Anderson of the sliding phason mode contribution to the conductivity 
in clean systems representing  the  perfect  conductivity  of  Fr\"ohlich superconductivity.  
In  the  presence  of impurities, which is always the case, phasons are pinned resulting 
in vanishing static conductivity at absolute zero (Appendix F).}

\ogata{So far we have reviewed in detail the derivation of phason contributions to $L_{11}$ 
in order to make transparent and solid the new contributions of phason drag to $L_{12}$ 
on equal footing to be explained in the following.}

\section{Thermoelectric conductivity due to phason drag}

In this section, we study the phason drag contribution to the thermoelectric conductivity, $L_{12}^{\rm ph}$, which is given by
\begin{equation}
L_{12}^{\rm ph} = \lim_{\omega\rightarrow 0} \frac{1}{i\omega} \left[ \Phi_{12}^{\rm ph} (i\omega_\lambda\rightarrow \hbar\omega+i\delta)
-\Phi_{12}^{\rm ph} (0)\right],
\end{equation}
with
\begin{equation}
\Phi_{12}^{\rm ph} (i\omega_\lambda) = \frac{1}{L} \int_0^\beta d\tau \langle 
T_\tau [ J_{\rm h}^{\rm ph}(\tau) J_e(0)] \rangle e^{i\omega_\lambda \tau},
\label{PhasonDCorr}
\end{equation}
and 
$J_{\rm h}^{\rm ph}$ is the heat current carried by phonon,\cite{OFSB}
\begin{equation}
J_{\rm h}^{\rm ph} = \sum_q \hbar \omega_q c_q b^\dagger_q b^{\phantom{\dagger}}_q, 
\label{PhononHeatC}
\end{equation}
with $c_q = d\omega_q /dq$ being the phonon group velocity.

\subsection{Phason drag process}

As in the case of FeSb$_2$,\cite{Matsuura} processes associated with phason drag are shown diagrammatically in Appendix C. 
Here it is to be noted that phonon propagators appearing in the phason drag processes are \lq\lq directed" 
as fermions,\cite{OFSB,Matsuura,Baumann,Sakuma,Okuma,Kohno,Kohno2} 
i.e., instead of ${\mathcal D}_{mn}(q,\tau)$, we have to use ${\mathcal O}_{mn}(q,\tau)$ defined as
\begin{equation}
{\mathcal O}_{mn}(q,\tau) = -\langle T_\tau 
[b^{\phantom{\dagger}}_{mQ+q} (\tau) \left( b^\dagger_{nQ+q} (0) + b^{\phantom{\dagger}}_{-nQ-q} (0)\right) ] \rangle.
\end{equation}
Details of calculations are shown in Appendices B and C. 
Finally, we obtain
\begin{eqnarray}
\Phi_{12}^{\rm ph} (i\omega_\lambda) &=& e \hbar v_{\rm F} \omega_Q c_Q g_Q^2 \frac{(k_{\rm B}T)^2}{L^2} \sum_{k,q, n,\nu} 
\frac{i\omega_\nu + \hbar\omega_Q}{2\hbar \omega_Q} 
\frac{i\omega_\nu +i\omega_\lambda + \hbar\omega_Q}{2\hbar \omega_Q} 
\frac{1}{(i \varepsilon_n)^2-E_k^2} \frac{1}{(i \varepsilon_n + i\omega_\lambda)^2-E_k^2} \cr
&&\times 
\biggl\{ P(q, i\omega_\nu) A(q, i\omega_\nu + i\omega_\lambda) 
+A(q, i\omega_\nu) P(q, i\omega_\nu + i\omega_\lambda) \biggr\} \cr
&&\times \biggl[ ( f(q, i\omega_\nu+i\omega_\lambda)-f(-q, -i\omega_\nu) )  
\left\{i \varepsilon_n (i \varepsilon_n+i\omega_\lambda)+ \xi_k^2 - \Delta_0^2\right\} \cr
&&- ( g(q, i\omega_\nu+i\omega_\lambda) - g(-q, -i\omega_\nu) )  
\xi_k (2i \varepsilon_n+i\omega_\lambda) \biggr] + O((i\omega_\lambda)^2), 
\label{Phixx}
\end{eqnarray}
with 
\begin{eqnarray}
f(q, i\omega_\nu) &=& \frac{i\varepsilon_n+i\omega_\nu}{(i\varepsilon_n + i\omega_\nu)^2-E_{k+q}^2}, \cr
g(q, i\omega_\nu) &=& \frac{\xi_{k+q}}{(i\varepsilon_n + i\omega_\nu)^2-E_{k+q}^2}.
\label{fgfunctDef}
\end{eqnarray}

From eq.~(\ref{Dyson}) we see that ${\mathcal D}_{mn}(-q, -i\omega_\nu) = {\mathcal D}_{mn}(q, i\omega_\nu)$, i.e., 
 $P(-q, -i\omega_\nu) = P(q, i\omega_\nu)$ and $A(-q, -i\omega_\nu) = A(q, i\omega_\nu)$. 
Thus changing variables, $q\rightarrow -q$ and $i\omega_\nu \rightarrow -i\omega_\nu - i\omega_\lambda$,
in $f(-q, -i\omega_\nu)$ and $g(-q, -i\omega_\nu)$ in eq.~(\ref{Phixx}), 
we obtain
\begin{eqnarray}
&&\Phi_{12}^{\rm ph} (i\omega_\lambda) = e \hbar v_{\rm F} \omega_Q c_Q g_Q^2 \frac{(k_{\rm B}T)^2}{L^2} \sum_{k,q, n,\nu} 
\frac{2i\omega_\nu +i\omega_\lambda}{2\hbar \omega_Q} 
\frac{1}{(i \varepsilon_n)^2-E_k^2} \frac{1}{(i\varepsilon_n + i\omega_\lambda )^2-E_k^2} \cr
&&\times 
\biggl\{ P(q, i\omega_\nu) A(q, i\omega_\nu + i\omega_\lambda) 
+A(q, i\omega_\nu) P(q, i\omega_\nu + i\omega_\lambda) \biggr\} \cr
&&\times \biggl[ f(q, i\omega_\nu+i\omega_\lambda) 
\left\{i \varepsilon_n (i \varepsilon_n+i\omega_\lambda)+ \xi_k^2 - \Delta_0^2\right\} 
- g(q, i\omega_\nu+i\omega_\lambda) 
\xi_k (2i \varepsilon_n+i\omega_\lambda) \biggr] + O((i\omega_\lambda)^2).
\label{Final}
\end{eqnarray}

\subsection{Analytic continuation and low temperature properties of $\Phi^{\rm ph}_{12}$}

For the static thermoelectric conductivity, we need to calculate the linear order of $i\omega_\lambda$
of eq.~(\ref{Final}), whose leading contributions are due to the region of $-\omega_\lambda<\omega_\nu<0$. 
[The other regions give contributions proportional to $P(q,x+i\delta) A(q,x+i\delta)$ or $P(q,x-i\delta) A(q,x-i\delta)$, 
which will be in the higher order with respect to damping of phason and amplitude mode.] 
After analytic continuation of $i\omega_\lambda \rightarrow \hbar \omega+i\delta$, we obtain
\begin{eqnarray}
\Phi_{12}^{\rm ph} (\hbar\omega+i\delta) &=& e \hbar v_{\rm F} \omega c_Q g_Q^2 \frac{1}{L} \sum_{q}
\int_{-\infty}^\infty \frac{dx}{2\pi i} n'(x) x\cr
&&\times \biggl\{ P(q, x-i\delta) A(q, x+i\delta)+A(q, x-i\delta) P(q, x+i\delta) \biggr\} C(x,q) + O(\omega^2),
\label{PhiFinal123}
\end{eqnarray}
where
\begin{eqnarray}
C(x,q) &=& \frac{k_{\rm B}T}{L} \sum_{k,n} \frac{1}{[(i \varepsilon_n)^2-E_k^2]^2} 
\frac{1}{(i \varepsilon_n+x)^2-E_{k+q}^2} \cr
&&\times 
\biggl\{ (i \varepsilon_n+x) 
\left\{ (i \varepsilon_n)^2+ \xi_k^2 - \Delta_0^2 \right\} - 2i \varepsilon_n \xi_k \xi_{k+q} \biggr\}.
\label{Cdef}
\end{eqnarray}
By noting that $C(x,q)$ is due to fermionic contributions with energy and momenta higher 
than those of phonons, we expand $C(x,q)$ in terms of both $x$ and $q$. 
In the lowest order with respect to $q$,
\begin{eqnarray}
C(x,q)\sim C(x,0) &=& \frac{k_{\rm B}T}{L} \sum_{k,n} \frac{1}{[(i \varepsilon_n)^2-E_k^2]^2}
\frac{1}{(i \varepsilon_n +x)^2 - E_k^2} \cr
&&\times 
\biggl\{ (i \varepsilon_n+x) 
\left\{ (i \varepsilon_n)^2+ \xi_k^2 - \Delta_0^2 \right\} - 2i \varepsilon_n \xi_k^2 \biggr\}.
 \label{CxqLowest0}
\end{eqnarray}
Since $n'(x)x$ is an odd function of $x$, 
the lowest order contributing to eq.~(\ref{PhiFinal123}) is $C(x,0) \sim xD(T)$ with
\begin{equation}
D(T) = - \frac{k_{\rm B}T}{L} \sum_{k,n} 
\frac{(i \varepsilon_n)^2- \xi_k^2 +\Delta_0^2}{[(i \varepsilon_n)^2-E_k^2]^3}.
\label{CxqLowest}
\end{equation}

Equation (\ref{CxqLowest}) is derived also rather straightforwardly by putting $q=0$ and $i\omega_\lambda=0$
in the electron Green's functions in Fig.~\ref{f1} (see Appendix D). 
Finally, the static thermoelectric conductivity at low temperature $L_{12}^{\rm ph} (T)$ is given as follows
\begin{eqnarray}
L_{12}^{\rm ph} (T) &=& -e \hbar v_{\rm F} c_Q g_Q^2 \frac{D(T)}{L} \sum_{q}
\int_{-\infty}^\infty \frac{dx}{2\pi} n'(x) x^2\cr
&&\times \biggl\{ P(q, x-i\delta) A(q, x+i\delta)+A(q, x-i\delta) P(q, x+i\delta) \biggr\}.
\label{L12Fin}
\end{eqnarray}
At $T=0$, 
$D(0) ={{k_{\rm F}}}/{8\pi E_{k_{\rm F}}^3}$, 
which is shown in Appendix E. 
%
We see that $L_{12}$ is governed by both phase and amplitude modes, while $L_{11}$ by phase mode only. 
Although impurity scattering affects both modes, phasons are more sensitive, which has been 
studied before in the context of impurity pinning which will be briefly summarized in the following.

\subsection{General features of phason propagators} 

In order to explore the implication of eq.~(\ref{L12Fin}), we analyze the propagators of collective modes 
of phason and amplitude, $P(q, \hbar\omega+i\delta)$ and $A(q, \hbar\omega+i\delta)$, 
given by eq.~(\ref{DefPhasonPropXX}) with TTF-TCNQ in mind 
\ogata{in the impurity-pinned state, i.e., in charge density glass (CDG) state instead of CDW state. 
We note that phonon propagators in glasses are proposed to be of the following type, e.g., in Ref.~\cite{Baggioli}}
\begin{eqnarray}
P(q, \hbar\omega+i\delta) &=& \frac{2\hbar\omega_Q/(1+X)}{(\hbar\omega+i\delta)^2 -\hbar^2 \omega_{\rm ph}^2(q) + i\hbar\omega \Gamma_{\rm ph}}, \cr
A(q, \hbar\omega+i\delta) &=& \frac{2\hbar\omega_Q/(1+X/3)}{(\hbar\omega+i\delta)^2 -\hbar^2 \omega_{\rm am}^2(q) + i\hbar\omega \Gamma_{\rm am}}, 
\label{PhasonandAmp}
\end{eqnarray}
where $\omega_{\rm ph}(q)=v |q|$ and $\omega_{\rm am}(q)$ are the dispersions of the phason and the amplitude mode, respectively. 
$\Gamma_{\rm ph}$ and $\Gamma_{\rm am}$ reflect the effects of randomness. 
\ogata{This expectation is justified for the amplitude mode which is optical and has a finite gap at $q=0$. 
However this expectation is totally invalid for phason, which is acoustic. } 
In the following we will see that $P(q, \hbar\omega+i\delta)$ is greatly modified because of the impurity pinning. 

We first note that these modes derived by the mean-field theory are to be valid in the three-dimensionally 
ordered Peierls phase. 
The critical temperature to the ordered Peierls phase is $T_{\rm P}\sim 54$K, 
which is believed to be much lower than the mean-field transition temperature 
$T_{{\rm P}0} \sim 500$K\cite{AndersonLeeSaitoh}
because of strong fluctuations intrinsic to one dimensionality. 
The wave-number $q$ is measured relative to 
$2k_{\rm F}$, since these are phonon modes in the Peierls phase with the long-range order parameter 
of coherent lattice distortion with period $2k_{\rm F}=Q$. 
As clarified by LRA, phasons carry charge current, while amplitude modes are neutral. 
This implies that phasons are considered to be charged phonons. 
Hence the present phasons have particular features compared with ordinary phonons: 
very low energy and sensitive to spatial randomness because of charged object. 

The subtle problem of the coupling of phasons to spatial randomness leading to impurity pinning 
had been studied before based on the effective Hamiltonian, phase Hamiltonian\cite{Fukuyama76,FukuLee}, 
which indicates that $P(q,\hbar\omega+i\delta)$ (\ref{PhasonandAmp}) at absolute zero is modified as follows,
\begin{equation}
P(q, \hbar\omega+i\delta) = \frac{2\hbar\omega_Q/(1+X)}{(\hbar\omega+i\delta)^2 -\hbar^2 v^2 q^2 - g_0 + i\hbar\omega g_1},
\label{PhasonandAmp2}
\end{equation}
where $g_0$ ($\sim \gamma^2$) and $g_1$ ($\sim \gamma$) are parameters associated with impurity pinning potential $\gamma$ ($\gamma>0$)
(for details, see Appendix F). 

By eq.~(\ref{PhasonandAmp2}) with finite $g_0$, it is seen that $\sigma(\omega) \sim i\omega$ as $\omega\rightarrow 0$, 
which is the characteristic of 
dielectrics (insulators) with the dielectric constant, $\varepsilon(\omega) = 1+ 4\pi i \sigma(\omega)/\omega \sim 1/(g_0-i\hbar \omega g_1)$. 
This reflects the fact that Peierls lattice distortions are no longer uniform in the pinned CDW state 
and that the spatial charge density is disordered, i.e., glassy.
In such a glassy state, Charge Density Glass (CDG) state,\cite{Fukuyama78} 
the possible charge transport is either uniform oscillations of phasons 
within each domain or local variation of phase associated with domain walls described as solitons 
both of which needs finite excitation energy.
These are features of impurity pinning at $T=0$ for finite frequency $\omega\ne 0$.  

At finite temperature, $T\ne 0$, these low energy excitations are thermally excited resulting in 
small but finite conductivity, which implies $g_0=0$ with finite $g_1$ in eq.~(\ref{PhasonandAmp2}).  
There will be an interesting crossover from the zero-temperature value of $g_0$ to vanishing $g_0$ at finite temperature, 
and this is associated with the dielectric anomalies which have characteristic dependences on both frequency and temperature 
of dielectric constant in some family of molecular solids.\cite{energyLand}
But this issue is beyond the scope of the present paper.
In the following, we assume $g_0=0$ for $T>0$. 
In this case, static conductivity $\sigma_0$ is given by
\begin{equation}
\sigma_0= \left( \frac{e}{\pi} \right)^2 \frac{g_Q^2}{(1+X)\Delta_0^2}
\frac{\omega_Q}{g_1}
\end{equation}
In the present context of TTF-TCNQ experiments indicate more or less the activation type of temperature dependence 
of conductivity \cite{Cohen}
implying $g_1(T) \sim \gamma{\rm exp}(E_0/k_{\rm B}T)$ which we will assume in the following.

\subsection{The temperature dependences of $L_{12}^{\rm ph} (T)$}

In order to see the implication of (\ref{L12Fin}), we first note the dispersion of amplitude 
mode $\omega_{\rm am}(q)$ is relatively weak 
compared to that of phasons, we assume $\omega_{\rm am}(q)$ is a $q$-independent constant, $\omega_a$. 
Then, $q$-integration in eq,~(\ref{L12Fin}) is possible analytically leading to 
\begin{eqnarray}
F(x, T) &=& \frac{1}{L} \sum_q \biggl\{ P(q, x-i\delta) A(q, x+i\delta)+A(q, x-i\delta) P(q, x+i\delta) \biggr\} \cr
&=& 2{\rm Re} \int_{-\infty}^\infty \frac{dq}{2\pi}\left( \frac{2\hbar\omega_Q/(1+X)}{x^2-\hbar^2 v^2 q^2-i g_1 x}\right)
\left( \frac{2\hbar\omega_Q/(1+X/3)}{x^2-\hbar^2 \omega_{\rm am}^2+ix \Gamma_{\rm am}} \right) \cr
&=& -{\rm Re} \left[ \frac{4i \hbar \omega_Q^2/v(1+X)(1+X/3)}{(x^2-\hbar^2 \omega_a^2+ix \Gamma_{\rm am}) ({x^2-ig_1 x})^{1/2}} 
\right],
\end{eqnarray}
where the argument of $({x^2-ig_1 x})^{1/2}$ is chosen as ${\rm Im}({x^2 -ig_1 x})^{1/2}>0$. 
Therefore, $L_{12}^{\rm ph} (T)$ in eq.~(\ref{L12Fin}) becomes
\begin{equation}
L_{12}^{\rm ph} (T) = -e \hbar^2 v_{\rm F}c_Q g_Q^2 D(T) \int_{-\infty}^\infty \frac{dx}{2\pi} n'(x) x^2 F(x,T).
\end{equation}
It is to be noted that the factor $|n'(x)|$ is large only for $|x| \lesssim k_{\rm B}T$ at low temperatures. 

As discussed in the previous subsection, when the system is conductive we expect 
$g_1\sim \gamma {\rm exp}(E_0/k_{\rm B}T)$. 
In the low temperatures where $g_1>> k_{\rm B}T$ holds, $F(x,T)$ is approximated as
\begin{equation}
F(x,T) = \frac{4 \omega_Q^2}{\sqrt{2}\hbar v \omega_a^2 (1+X)(1+X/3)}\frac{1}{\sqrt{g_1(T) |x|}},
\end{equation}
and then
\begin{eqnarray}
L_{12}^{\rm ph} (T) &=& \frac{e \hbar v_{\rm F} c_Q g_Q^2\omega_Q^2 }{\sqrt{2} \pi v \omega_a^2 (1+X)(1+X/3)} 
\frac{(k_{\rm B}T)^{3/2}}{\sqrt{g_1(T)}}D(T)
\int_{0}^\infty dz \frac{z^{3/2}}{\sinh^2 z} \cr
&=& 2.936 \frac{e \hbar v_{\rm F} c_Q g_Q^2\omega_Q^2 }{\sqrt{2} \pi v \omega_a^2 (1+X)(1+X/3)} 
\frac{(k_{\rm B}T)^{3/2}}{\sqrt{g_1(T)}}D(T)
\end{eqnarray}
which lead to 
\begin{equation}
\ogata{
S = \frac{L_{12}^{\rm ph} }{T\sigma} 
\propto v_{\rm F} c_Q T^{1/2} e^{E_0/2k_{\rm B}T}.}
\end{equation}
\ogata{Here $D(T)$ has been approximated as a constant, $D(0)$. }

It is to be noted that $|S|$ is exponentially diverging toward absolute zero 
in the present 1d Peierls model, where the energy dispersion of electronic band is strictly 1d. 
The sign of $S$ is determined by $c_Q=d\omega_q/dq |_{q=Q}$ ($Q=2k_{\rm F}$) 
since $v_{\rm F}>0$ is independent of the filling of the band in the 
present 1d electron model: 
$S>0$ for $0<Q<G/2$ (\lq\lq electrons") and 
$S<0$ for $G/2<Q<G$ (\lq\lq holes") with $G$ being the reciprocal lattice vectors. 
In the case of charge transfer salts of our interest, TTF-TCNQ, however, system is semimetallic 
with same number of electrons and holes in TCNQ band and TTF band, respectively, 
and then $v_{\rm F}>0$ for electrons and $v_{\rm F}<0$ for holes as in doped semiconductors.

Present results may point to an interesting possibility of thermoelectricity in disordered (glassy) 
systems with strong electron-phonon coupling between low temperature insulating and 
weakly conducting intermediate temperature regions which may include some cases of variable range hopping.

\section{Summary}

In the present paper, effects of phason drag on the Seebeck coefficient has been 
theoretically studied for the one-dimensional incommensurate Peierls phase with TTF-TCNQ in mind 
based on the Kubo-Luttinger formalism with the help of thermal Green function. 
The phason is the collective mode of electron-lattice coupled CDW (charge density wave) systems 
and represent the sliding motions of electronic charge density and lattice distortion 
as clarified by Lee, Rice and Anderson (LRA). 
Hence phason can be considered as the ultimate form of phonon drag, which has long 
been known to play important roles in semiconductors and also in FeSb$_2$ recently. 

In order to treat phason dynamics theoretically, it is crucial to note the existence of 
two energy scales, i.e., high energy region representing electronic degrees of freedom 
to support the Peierls phase and the low energy region describing the collective modes 
(amplitude and phase modes) in the Peierls ordered state. 
As demonstrated by LRA, phasons are charged while amplitude modes are neutral. 
Phasons which represent sliding motions of coupled electronic charge density and 
lattice distortions, have acoustic wave vector dependence and leads to perfect electric conduction 
(Fr\" ohlich superconductivity) in clean systems.
However phasons are sensitive to spatial inhomogeneity in contrast to the phase of 
superconductivity and easily pinned by impurities resulting in insulating state at absolute zero 
with inhomogeneous spatial charge density. i.e., charge density glass (CDG) state.
In order to describe this dramatic processes of pinning from perfect conduction to 
insulating CDG state the phase Hamiltonian, which is effective Hamiltonian focusing on phasons, 
is known to be powerful to see the frequency dependences of conductivity at $T=0$. 
In the present studies on Seebeck coefficient we need to extend this study to finite temperatures.

We first demonstrated the perfect correspondences between former diagrammatical calculations of conductivity, 
$L_{11}$, and thermoelectric conductivity, $L_{12}$, and those based on Phase Hamiltonian 
in the absence of pining. 
Then effects of pinning on phasons governing $L_{11}$ at finite temperatures have first been 
analyzed based on former analysis at absolute zero (but finite frequencies).
This partly corresponds to general studies on phonon propagators in disordered systems, i.e.,
phonons in glassy states. However there is an important difference between phonons 
in glassy state and present CDG state: 
phonons are neutral in the former while they are charged here. 
In CDG state the dependences on frequency and temperature of phasons are more subtle than in neutral phonons. 
With such detailed studies on phason propagators in CDG state, its drag effects on $L_{12}$ and 
then $S=L_{12}/TL_{11}$, have been identified. 
It turns out that $|S|$ can be very large:
When conductivity obeys the Arrhenius type of temperature dependence, $L_{11} \propto e^{-E_0/k_{\rm B}T}$, 
\ogata{
then $L_{12}^{\rm ph}\propto v_{\rm F} c_Q T^{3/2} e^{-E_0/2k_{\rm B}T}$, and 
$S\propto v_{\rm F} c_Q T^{1/2} e^{E_0/2k_{\rm B}T}$} as $T\rightarrow 0$.
The sign of \ogata{$S$} is always opposite to electronic contributions which appear to be 
consistent with experiments,\cite{Kwak} 
although the description of crossover regions between 
high temperature with electronic contributions and the present low temperatures deep 
in Peierls ordered state is beyond the scope of present paper.

\ogata{
The main result of this paper is the identification of the phason drag contribution to the thermoelectric conductivity 
$L_{12}$,  eq.~(\ref{L12Fin}),  in terms phason  and  amplitude  propagators, $P(q,x), A(q,x)$,  
to be combined with the conductivity, $L_{11}$,  eq.~(\ref{phasonSigma}),  for  the  Peierls  phase treated  
within  the  mean-field  theory. 
Even if the Peierls phase is treated in more detail beyond the mean-field theory, the main framework of 
the present scheme will be valid for  the  contribution  of  phase  and  amplitude  modes  
as  far  as  the  Peierls  phase  is long-range  ordered  and  stable with  possible  modifications  
of  the  prefactors  of  eqs.~(\ref{phasonSigma}) and (\ref{L12Fin}).}
\bigskip

{Acknowledgments}

This paper is dedicated to Professor Phil Anderson, who passed very recently 
while we had been preparing the manuscript, for his continual encouragement 
through enlightening discussion for many years
We thank very fruitful discussions with H.\ Matsuura and H.\ Maebashi. 
HF thanks Patrick Lee and Maurice Rice for useful discussions in March and September, 2019, respectively.
This work was supported by Grants-in-Aid for Scientific
Research from the Japan Society for the Promotion of Science (Grants No. JP18H01162), 
and by JST-Mirai Program Grant Number JPMJMI19A1, Japan.





\appendix

\section{Dyson equations for phonon propagators} 

\begin{figure}
\includegraphics[width=16cm]{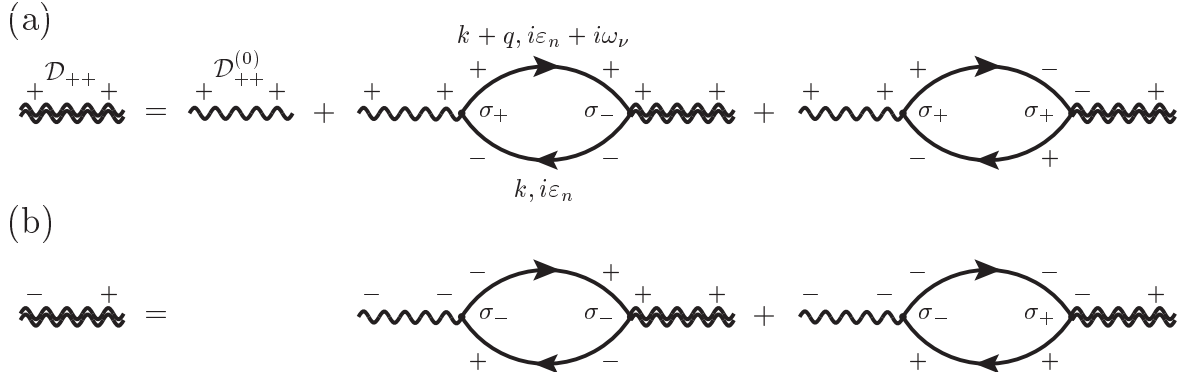} 
\caption{Feynman diagrams of Dyson equations for phonon propagator $\mathcal D_{mn}$.
The wavy line (the double wavy line) represents $\mathcal D^{(0)}_{mn}$ ($\mathcal D_{mn}$). 
The $\pm$ signs attached to the wavy lines and the solid lines 
represent the subscripts of the corresponding propagator. 
The Dyson equations for $\mathcal D_{--}$ and $\mathcal D_{-+}$ are written in the same way.
}
\label{f2}
\end{figure}

The Dyson equations for ${\mathcal D}_{mn}$ by LRA are shown in Fig.~\ref{f2} which lead
\begin{eqnarray}
{\mathcal D}_{++}(q, i\omega_\nu) &=& {\mathcal D}^{(0)}_{++}(q, i\omega_\nu)
+ {\mathcal D}_{++}(q, i\omega_\nu) {\Pi}_{++}(q, i\omega_\nu) {\mathcal D}^{(0)}_{++}(q, i\omega_\nu) \cr
&&+ {\mathcal D}_{+-}(q, i\omega_\nu) {\Pi}_{-+}(q, i\omega_\nu) {\mathcal D}^{(0)}_{++}(q, i\omega_\nu), \cr
{\mathcal D}_{+-}(q, i\omega_\nu) 
&=& {\mathcal D}_{++}(q, i\omega_\nu) {\Pi}_{+-}(q, i\omega_\nu) {\mathcal D}^{(0)}_{--}(q, i\omega_\nu) 
+ {\mathcal D}_{+-}(q, i\omega_\nu) {\Pi}_{--}(q, i\omega_\nu) {\mathcal D}^{(0)}_{--}(q, i\omega_\nu), \cr
{\mathcal D}_{-+}(q, i\omega_\nu) 
&=& {\mathcal D}_{--}(q, i\omega_\nu) {\Pi}_{-+}(q, i\omega_\nu) {\mathcal D}^{(0)}_{++}(q, i\omega_\nu) 
+ {\mathcal D}_{-+}(q, i\omega_\nu) {\Pi}_{++}(q, i\omega_\nu) {\mathcal D}^{(0)}_{++}(q, i\omega_\nu), \cr
{\mathcal D}_{--}(q, i\omega_\nu) &=& {\mathcal D}^{(0)}_{--}(q, i\omega_\nu)
+ {\mathcal D}_{--}(q, i\omega_\nu) {\Pi}_{--}(q, i\omega_\nu) {\mathcal D}^{(0)}_{--}(q, i\omega_\nu) \cr
&&+ {\mathcal D}_{-+}(q, i\omega_\nu) {\Pi}_{+-}(q, i\omega_\nu) {\mathcal D}^{(0)}_{--}(q, i\omega_\nu),
\label{AppA1}
\end{eqnarray}
where $\Pi_{mn}$ are defined in eq.~(\ref{PiDef}). 
It is to be noted that the relations $\Pi_{++}=\Pi_{--}$ and $\Pi_{-+}=e^{-4i\phi} \Pi_{+-}$ hold from their definitions. 
(Note that in the presence of $\overline \varepsilon_k$ the relation $\Pi_{++}=\Pi_{--}$ does not hold.) 
Furthermore, when we use an approximation [eq.~(\ref{D0Approxi})]
\begin{equation}
{\mathcal D}^{(0)}_{mn}(q,i\omega_\nu) \sim \delta_{mn} {\mathcal D}^{(0)}(i\omega_\nu) 
= \delta_{mn} \frac{2\hbar \omega_Q}{(i\omega_\nu)^2 - \hbar^2 \omega_Q^2},
\end{equation}
eq.~(\ref{AppA1}) becomes
\begin{eqnarray}
{\mathcal D}_{++}(q, i\omega_\nu) &=& \left\{ 
1 + {\mathcal D}_{++}(q, i\omega_\nu) {\Pi}_{++}(q, i\omega_\nu) 
+ {\mathcal D}_{+-}(q, i\omega_\nu) e^{-4i\phi} {\Pi}_{+-}(q, i\omega_\nu) \right\} {\mathcal D}^{(0)}(i\omega_\nu), \cr
{\mathcal D}_{+-}(q, i\omega_\nu) 
&=& \left\{ {\mathcal D}_{++}(q, i\omega_\nu) {\Pi}_{+-}(q, i\omega_\nu) 
+ {\mathcal D}_{+-}(q, i\omega_\nu) {\Pi}_{++}(q, i\omega_\nu)\right\} {\mathcal D}^{(0)}(i\omega_\nu), \cr
{\mathcal D}_{-+}(q, i\omega_\nu) 
&=& \left\{ {\mathcal D}_{--}(q, i\omega_\nu) e^{-4i\phi} {\Pi}_{+-}(q, i\omega_\nu) 
+ {\mathcal D}_{-+}(q, i\omega_\nu) {\Pi}_{++}(q, i\omega_\nu)\right\} {\mathcal D}^{(0)}(i\omega_\nu), \cr
{\mathcal D}_{--}(q, i\omega_\nu) &=& \left\{
1+ {\mathcal D}_{--}(q, i\omega_\nu) {\Pi}_{++}(q, i\omega_\nu) 
+ {\mathcal D}_{-+}(q, i\omega_\nu) {\Pi}_{+-}(q, i\omega_\nu)\right\} {\mathcal D}^{(0)}(i\omega_\nu).
\end{eqnarray}
From these Dyson equations, we can see that 
\begin{eqnarray}
{\mathcal D}_{++}(q, i\omega_\nu) \pm e^{-2i\phi} {\mathcal D}_{+-}(q, i\omega_\nu)  
&=& \biggl[
1 + \left\{ {\mathcal D}_{++}(q, i\omega_\nu) \pm e^{-2i\phi} {\mathcal D}_{+-}(q, i\omega_\nu)\right\} \cr
&&\times
\left\{ {\Pi}_{++}(q, i\omega_\nu) \pm e^{-2i\phi} {\Pi}_{+-}(q, i\omega_\nu)\right\} \biggr] {\mathcal D}^{(0)}(i\omega_\nu),
\label{AppA2}
\end{eqnarray}
which leads to eq.~(\ref{Dyson}).
In a similar way, we obtain ${\mathcal D}_{--}={\mathcal D}_{++}$ and ${\mathcal D}_{-+}=e^{-4i\phi} {\mathcal D}_{+-}$.

The phason propagator $P(q,i\omega_\nu)$ in eq.~(\ref{DefPhasonAmp}) is evaluated 
in the small-$q$ and small-$(i\omega_\nu)$ region as follows. 
Substituting the definition of ${\mathcal D}^{(0)}(q, i\omega_\nu)$, $P(q,i\omega_\nu)$ is rewritten as
\begin{equation}
P(q, i\omega_\nu) 
= \frac{2\hbar \omega_Q}{(i\omega_\nu)^2 - (\hbar\omega_Q)^2 -2\hbar\omega_Q 
\left[\Pi_{++}(q, i\omega_\nu) - e^{-2i\phi} \Pi_{+-}(q, i\omega_\nu) \right]}. \label{AppendixAPropPP}
\end{equation}
Using the definition in eq.~(\ref{PiDef}) and the Green's function in eq.~(\ref{GreenF}), we obtain
\begin{equation}
{\Pi}_{++}(q, i\omega_\nu) - e^{-2i\phi} {\Pi}_{+-}(q, i\omega_\nu) = 2g_Q^2 \frac{k_{\rm B}T}{L} \sum_{k,n}
\frac{(i\varepsilon_n+i\omega_\nu +\xi_{k+q}) (i\varepsilon_n-\xi_{k})-\Delta_0^2}
{[(i\varepsilon_n+i\omega_\nu)^2-E_{k+q}^2] [(i\varepsilon_n)^2-E_{k}^2]}. \label{PhasonDeno1}
\end{equation}
When $q=0$ and $i\omega_\nu=0$, the r.h.s.\ of eq.~(\ref{PhasonDeno1}) becomes
\begin{eqnarray}
2g_Q^2 \frac{k_{\rm B}T}{L} \sum_{k,n} \frac{1}{(i\varepsilon_n)^2-E_{k}^2} 
&=& -2g_Q^2 \frac{1}{L} \sum_{k} \oint \frac{dz}{2\pi i} \frac{f(z)}{z^2-E_{k}^2} \cr
&=& 2g_Q^2 \frac{1}{L} \sum_{k} \frac{f(E_k)-f(-E_k)}{2E_{k}} \cr
&=& -\frac{\hbar\omega_Q}{2}, 
\end{eqnarray}
where the self-consistency equation in eq.~(\ref{SCeqForD}) has been used. 
The phason velocity is obtained by calculating the higher-order terms with respect to $q$ and $i\omega_\nu$. 
It is straightforward to obtain 
\begin{equation}
{\Pi}_{++}(q, i\omega_\nu) - e^{-2i\phi} {\Pi}_{+-}(q, i\omega_\nu) = 
-\frac{\hbar\omega_Q}{2} + \frac{X(\hbar v_{\rm F} q)^2}{2\hbar\omega_Q} 
- \frac{X (i\omega_\nu)^2}{2\hbar \omega_Q} + ({\rm higher}\ {\rm order} \ {\rm terms}).
\label{PhasonDeno2}
\end{equation}
with 
\begin{equation}
X=\frac{\omega_Q g_Q^2}{2\pi |v_{\rm F}| \Delta_0^2}.
\end{equation}
Substituting (\ref{PhasonDeno2}) into (\ref{AppendixAPropPP}), the phason propagator becomes
\begin{eqnarray}
P(q, i\omega_\nu) 
&=& \frac{2\hbar \omega_Q}{(i\omega_\nu)^2 -X (\hbar v_{\rm F} q)^2 + X(i\omega_\nu)^2} \cr
&=& \frac{2\hbar \omega_Q/(1+X)}{(i\omega_\nu)^2 - \frac{X}{1+X} (\hbar v_{\rm F} q)^2} \cr
&=& \frac{2\hbar \omega_Q/(1+X)}{(i\omega_\nu)^2 - (\hbar v q)^2}, 
\end{eqnarray}
where $v$ represents the phason velocity defined as $v=(X/(1+X))^{1/2}|v_{\rm F}|$. 
In the similar way, we obtain
\begin{eqnarray}
{\Pi}_{++}(q, i\omega_\nu) + e^{-2i\phi} {\Pi}_{+-}(q, i\omega_\nu) &=& 
-\frac{\hbar\omega_Q}{2} + \frac{2X\Delta_0^2}{\hbar\omega_Q} \cr
&&+ \frac{X(\hbar v_{\rm F} q)^2}{6\hbar\omega_Q}  
- \frac{X (i\omega_\nu)^2}{6\hbar \omega_Q} + ({\rm higher}\ {\rm order} \ {\rm terms}).
\label{AmpliDeno2}
\end{eqnarray}
Therefore, the amplitude propagator becomes
\begin{eqnarray}
A(q, i\omega_\nu) 
&=& \frac{2\hbar \omega_Q}{(i\omega_\nu)^2 - 4X \Delta_0^2- \frac{X}{3} (\hbar v_{\rm F} q)^2 + \frac{X}{3}(i\omega_\nu)^2} \cr
&=& \frac{2\hbar \omega_Q/(1+X/3)}{(i\omega_\nu)^2 - \hbar^2 \omega_{\rm am}^2(q)},
\end{eqnarray}
with 
\begin{equation}
\omega_{\rm am}(q) = \sqrt{\frac{4X}{1+\frac{X}{3}}\Delta_0^2 + \frac{X}{3+X}(\hbar v_{\rm F}q)^2}.
\end{equation}

\section{Dyson equations for \lq\lq directed" phonon propagators}

\begin{figure}
\includegraphics[width=16cm]{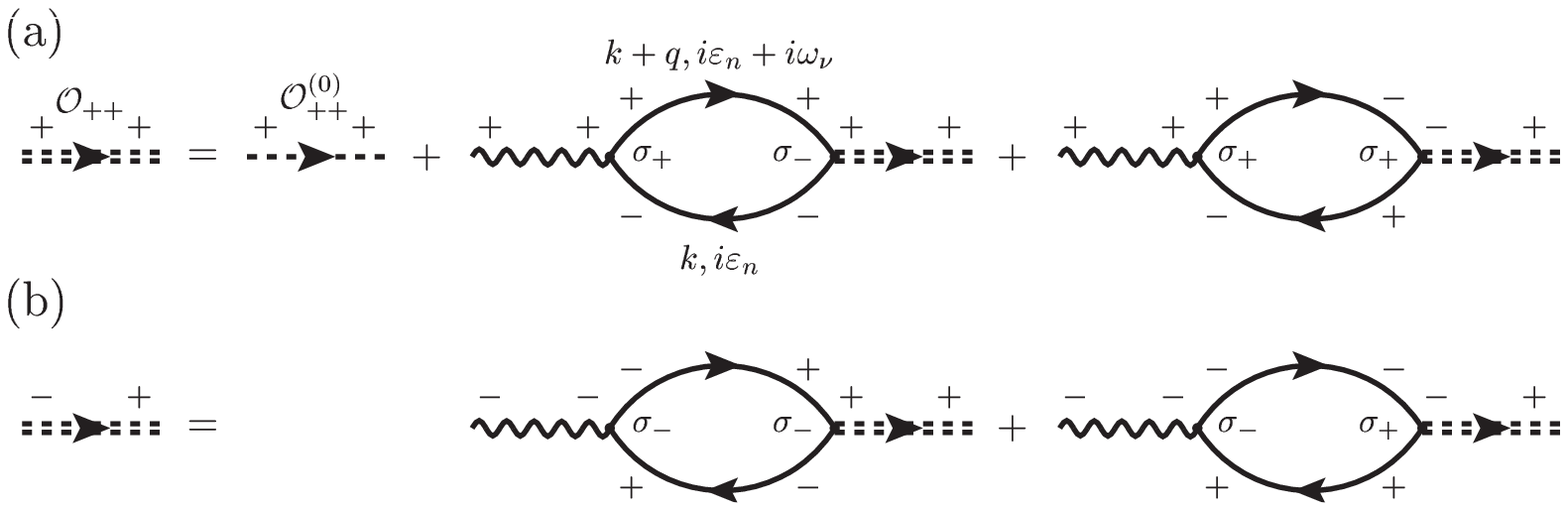} 
\caption{Feynman diagrams of Dyson equations for the \lq\lq directed" phonon propagator $\mathcal O_{mn}$.
The dashed line (the double dashed line) with an arrow represents $\mathcal O^{(0)}_{mn}$ ($\mathcal O_{mn}$). 
The $\pm$ signs attached to the dashed lines and the solid lines 
represent the subscripts of the corresponding propagator. 
The Dyson equations for $\mathcal O_{--}$ and $\mathcal O_{-+}$ are written in the same way.
}
\label{f3}
\end{figure}

The Dyson equations for $\mathcal O_{mn}$ are shown in Fig.~\ref{f3} which lead
\begin{eqnarray}
{\mathcal O}_{++}(q, i\omega_\nu) &=& {\mathcal O}^{(0)}_{++}(q, i\omega_\nu)
+ {\mathcal O}_{++}(q, i\omega_\nu) {\Pi}_{++}(q, i\omega_\nu) {\mathcal D}^{(0)}_{++}(q, i\omega_\nu) \cr
&&+ {\mathcal O}_{+-}(q, i\omega_\nu) {\Pi}_{-+}(q, i\omega_\nu) {\mathcal D}^{(0)}_{++}(q, i\omega_\nu), \cr
{\mathcal O}_{+-}(q, i\omega_\nu) 
&=& {\mathcal O}_{++}(q, i\omega_\nu) {\Pi}_{+-}(q, i\omega_\nu) {\mathcal D}^{(0)}_{--}(q, i\omega_\nu) 
+ {\mathcal O}_{+-}(q, i\omega_\nu) {\Pi}_{--}(q, i\omega_\nu) {\mathcal D}^{(0)}_{--}(q, i\omega_\nu), \cr
{\mathcal O}_{-+}(q, i\omega_\nu) 
&=& {\mathcal O}_{--}(q, i\omega_\nu) {\Pi}_{-+}(q, i\omega_\nu) {\mathcal D}^{(0)}_{++}(q, i\omega_\nu) 
+ {\mathcal O}_{-+}(q, i\omega_\nu) {\Pi}_{++}(q, i\omega_\nu) {\mathcal D}^{(0)}_{++}(q, i\omega_\nu), \cr
{\mathcal O}_{--}(q, i\omega_\nu) &=& {\mathcal O}^{(0)}_{--}(q, i\omega_\nu)
+ {\mathcal O}_{--}(q, i\omega_\nu) {\Pi}_{--}(q, i\omega_\nu) {\mathcal D}^{(0)}_{--}(q, i\omega_\nu) \cr
&&+ {\mathcal O}_{-+}(q, i\omega_\nu) {\Pi}_{+-}(q, i\omega_\nu) {\mathcal D}^{(0)}_{--}(q, i\omega_\nu), 
\end{eqnarray}
where ${\mathcal O}^{(0)}_{mn} (q, i\omega_\nu)$ are defined as
\begin{equation}
{\mathcal O}^{(0)}_{mn} (q, i\omega_\nu) = \frac{\delta_{mn}}{i\omega_\nu - \hbar \omega_{mQ+q}}. 
\end{equation}

Solving the Dyson equation for ${\mathcal O}_{mn}$, we obtain simple relations
\begin{equation}
{\mathcal O}_{mn}(q, i\omega_\nu) 
=\frac{{\mathcal O}^{(0)}(i\omega_\nu)}{{\mathcal D}^{(0)}(i\omega_\nu)} {\mathcal D}_{mn}(q, i\omega_\nu) 
=\frac{i\omega_\nu + \hbar\omega_Q}{2\hbar \omega_Q} {\mathcal D}_{mn}(q, i\omega_\nu).
\label{SimpleRel}
\end{equation}
where it is to be noted that we used an approximation 
\begin{equation}
{\mathcal O}^{(0)}_{mn}(q,i\omega_\nu) \sim \delta_{mn} {\mathcal O}^{(0)}(i\omega_\nu) 
= \delta_{mn} \frac{1}{i\omega_\nu- \hbar\omega_Q},
\end{equation}
as for ${\mathcal D}^{(0)}_{mn}(q,i\omega_\nu)$. 
The same argument is applied to $\tilde {\mathcal O}_{mn}(q, i\omega_\nu)$ which lead to
$\tilde{\mathcal O}_{mn}(q, i\omega_\nu) = {\mathcal O}_{mn}(q, i\omega_\nu)$.

\section{Feynman diagrams for the phason drag}

\begin{figure}
\includegraphics[width=6cm]{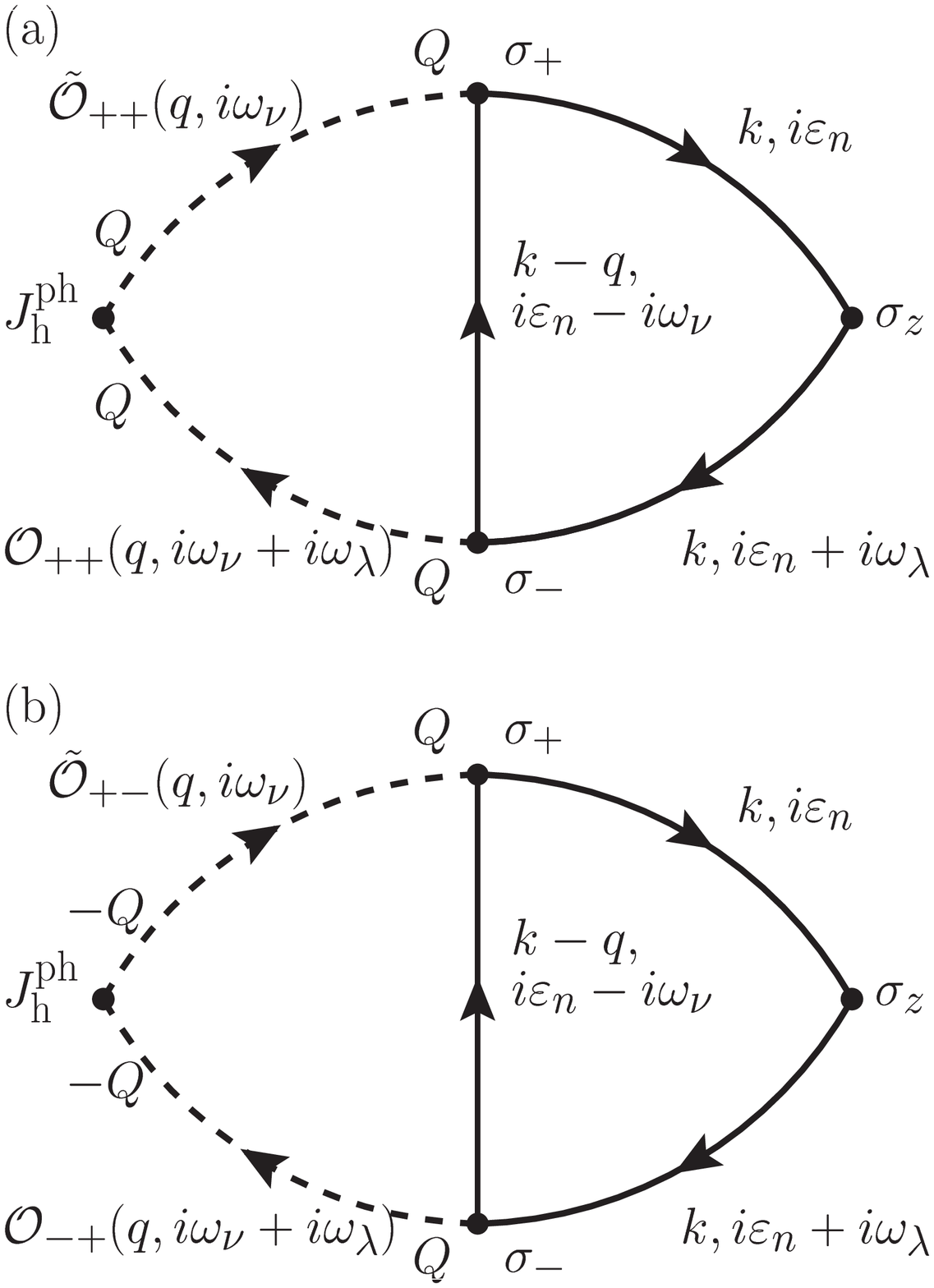} \hskip 0.5 cm
\includegraphics[width=6cm]{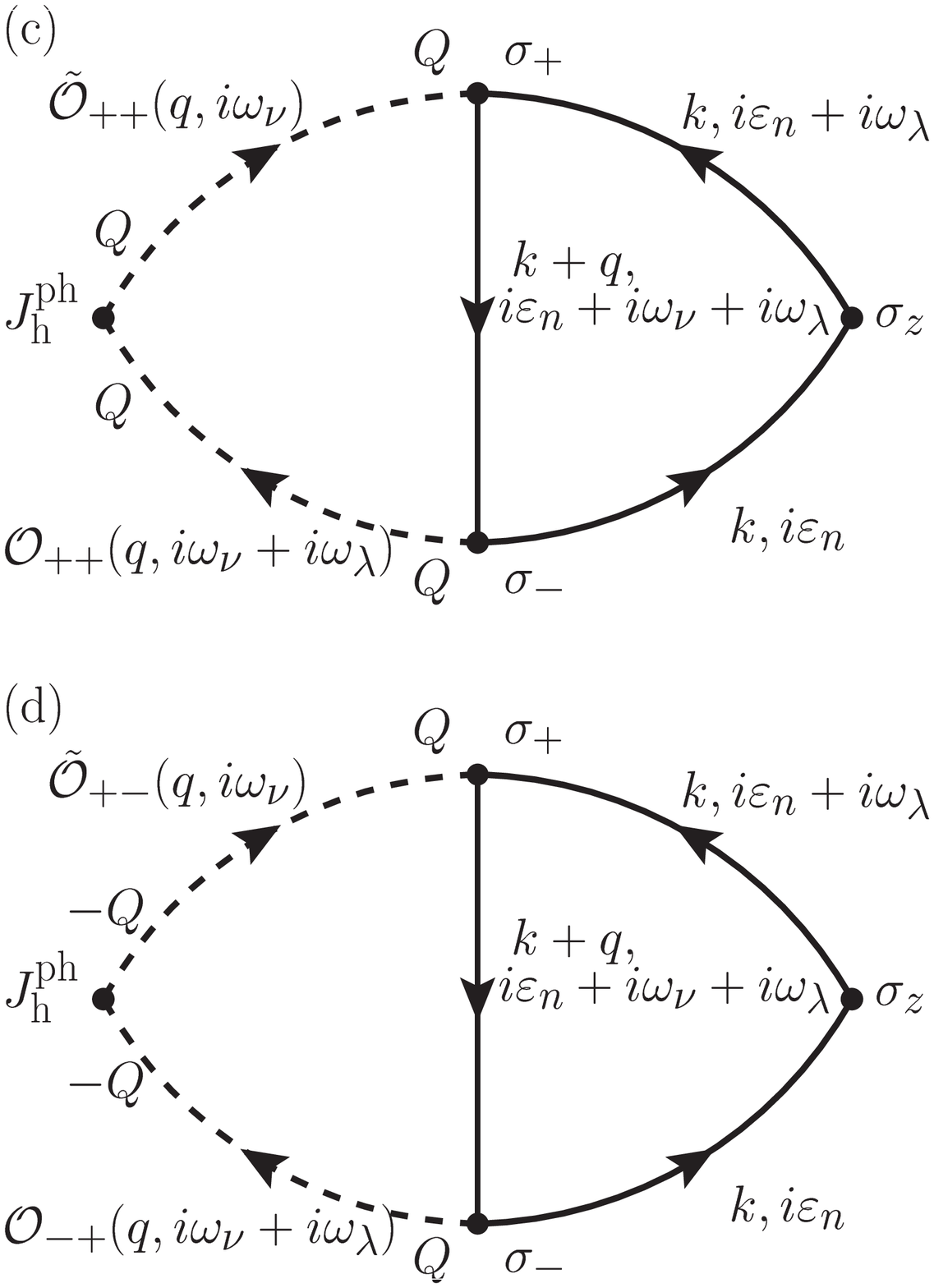} 
\includegraphics[width=4.5cm]{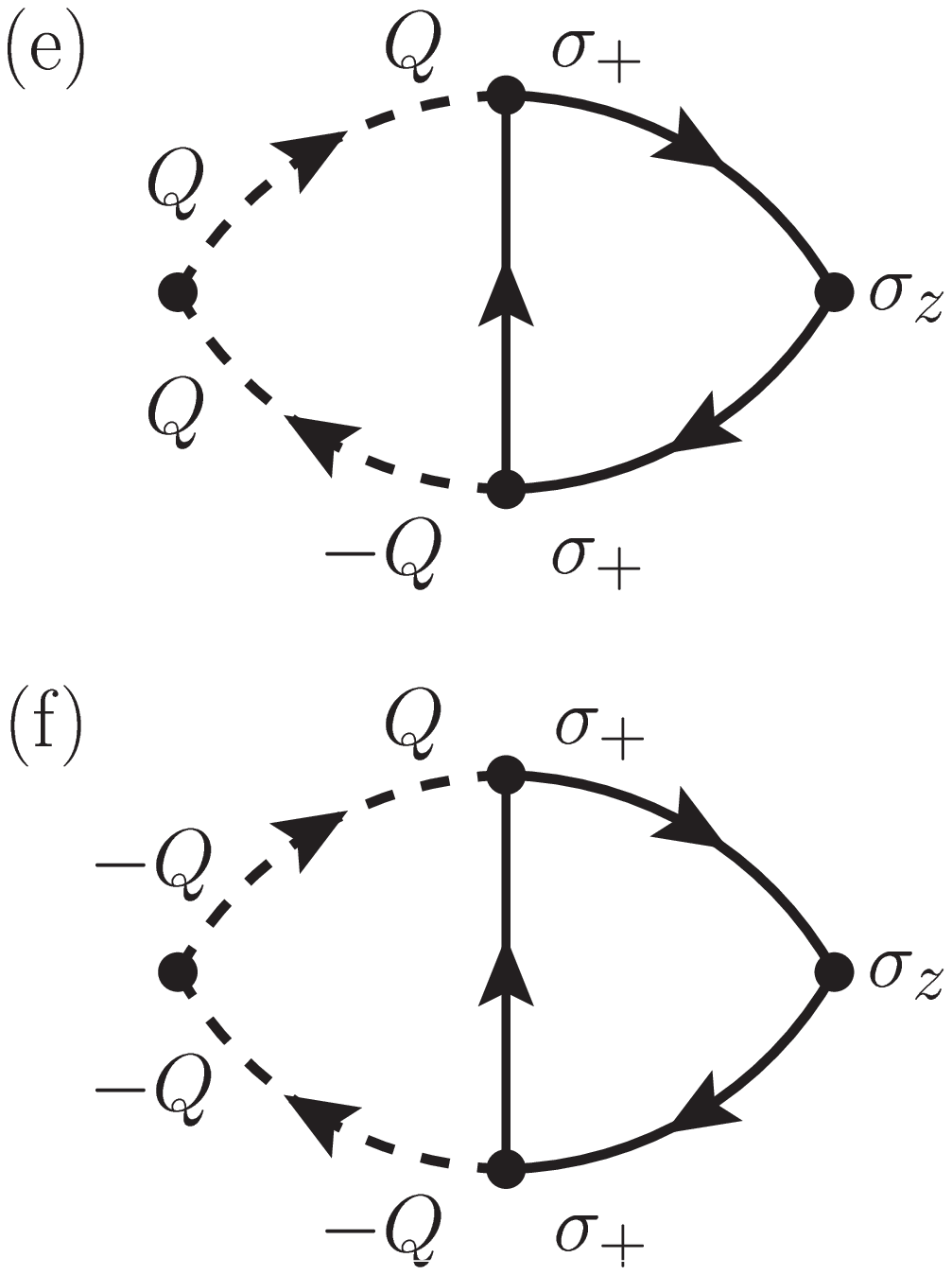}
\includegraphics[width=4.5cm]{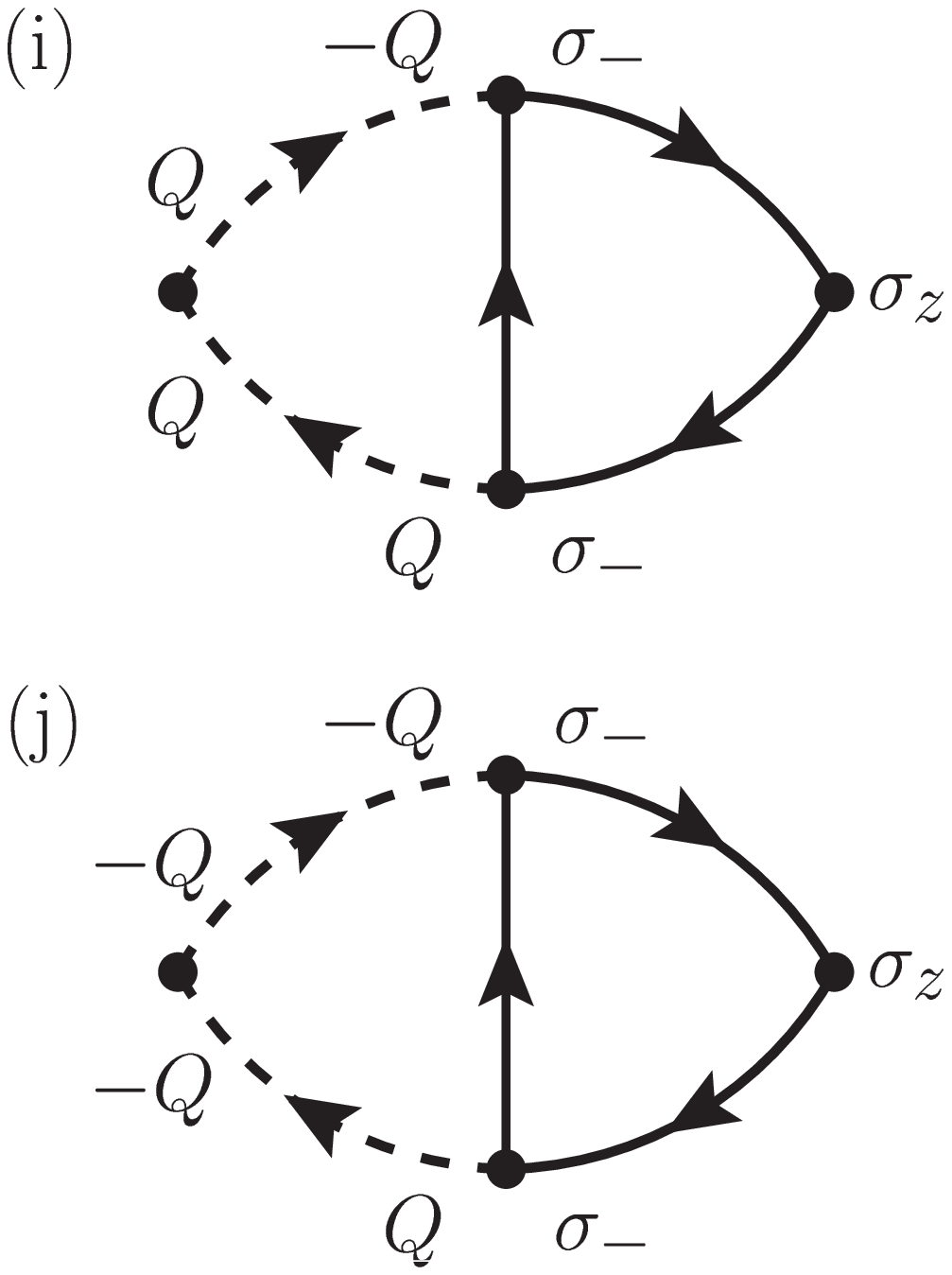} 
\includegraphics[width=4.5cm]{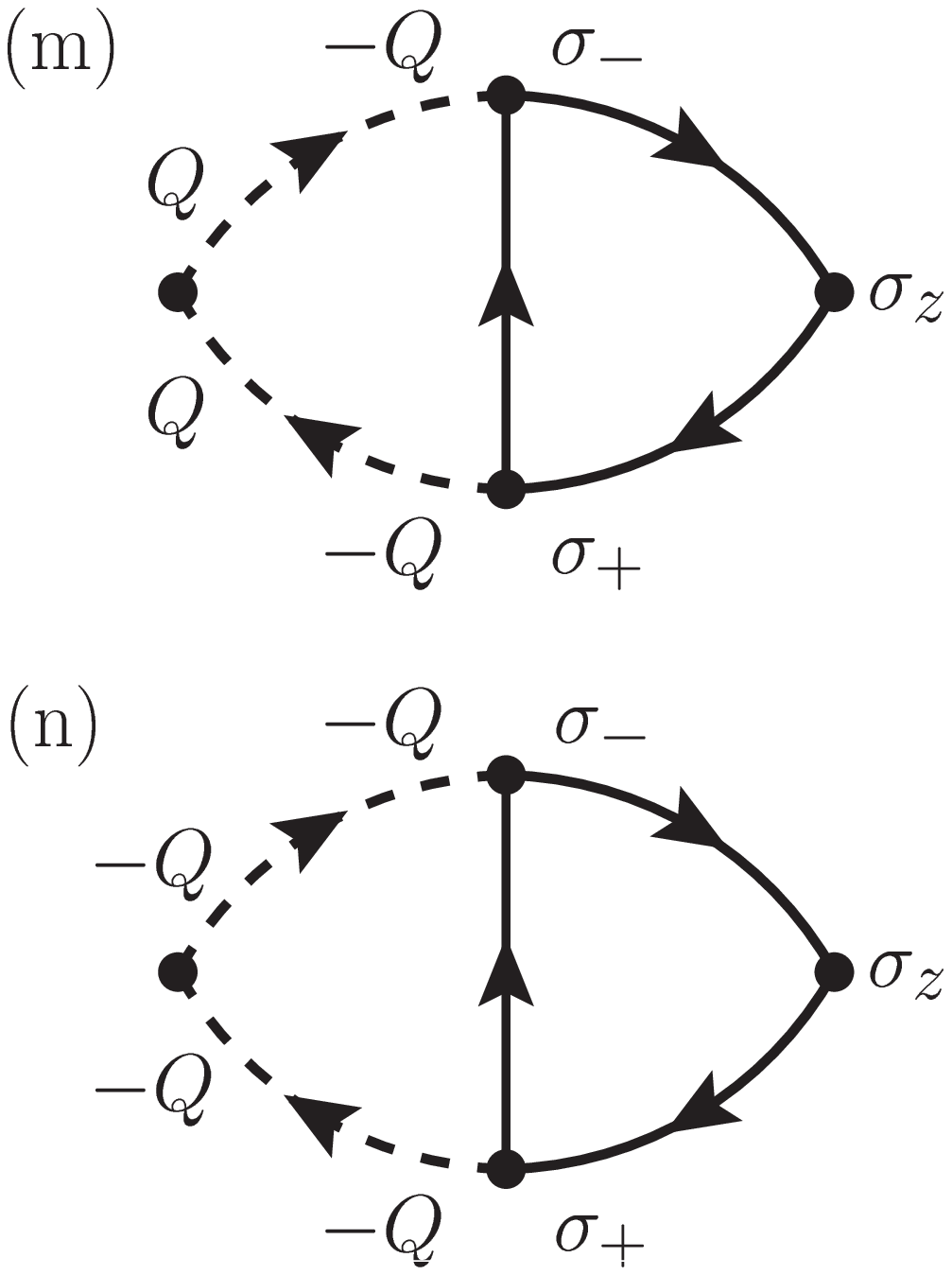} 
\includegraphics[width=4.5cm]{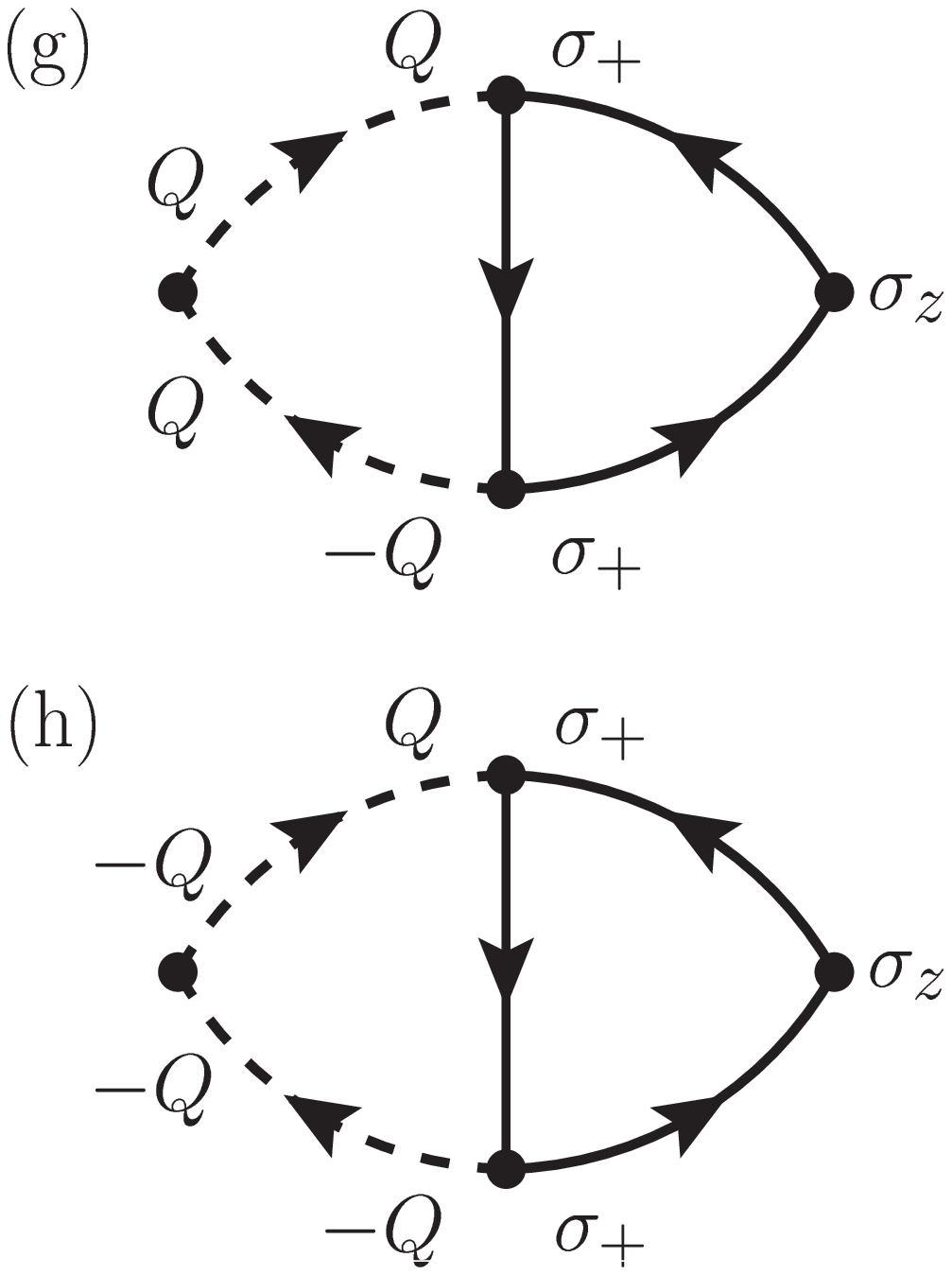} \hskip 1 cm
\includegraphics[width=4.5cm]{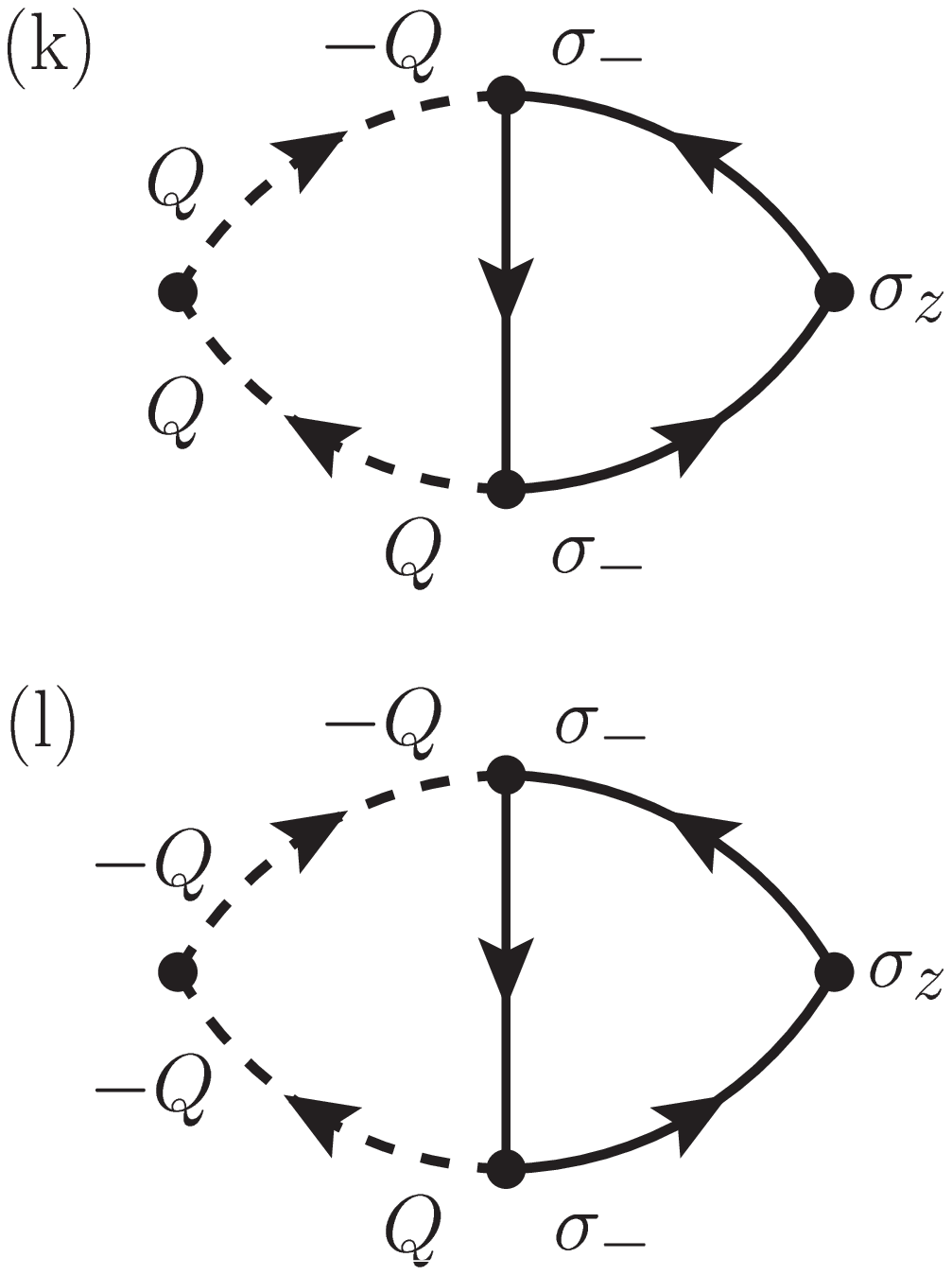} \hskip 0.8 cm
\includegraphics[width=4.5cm]{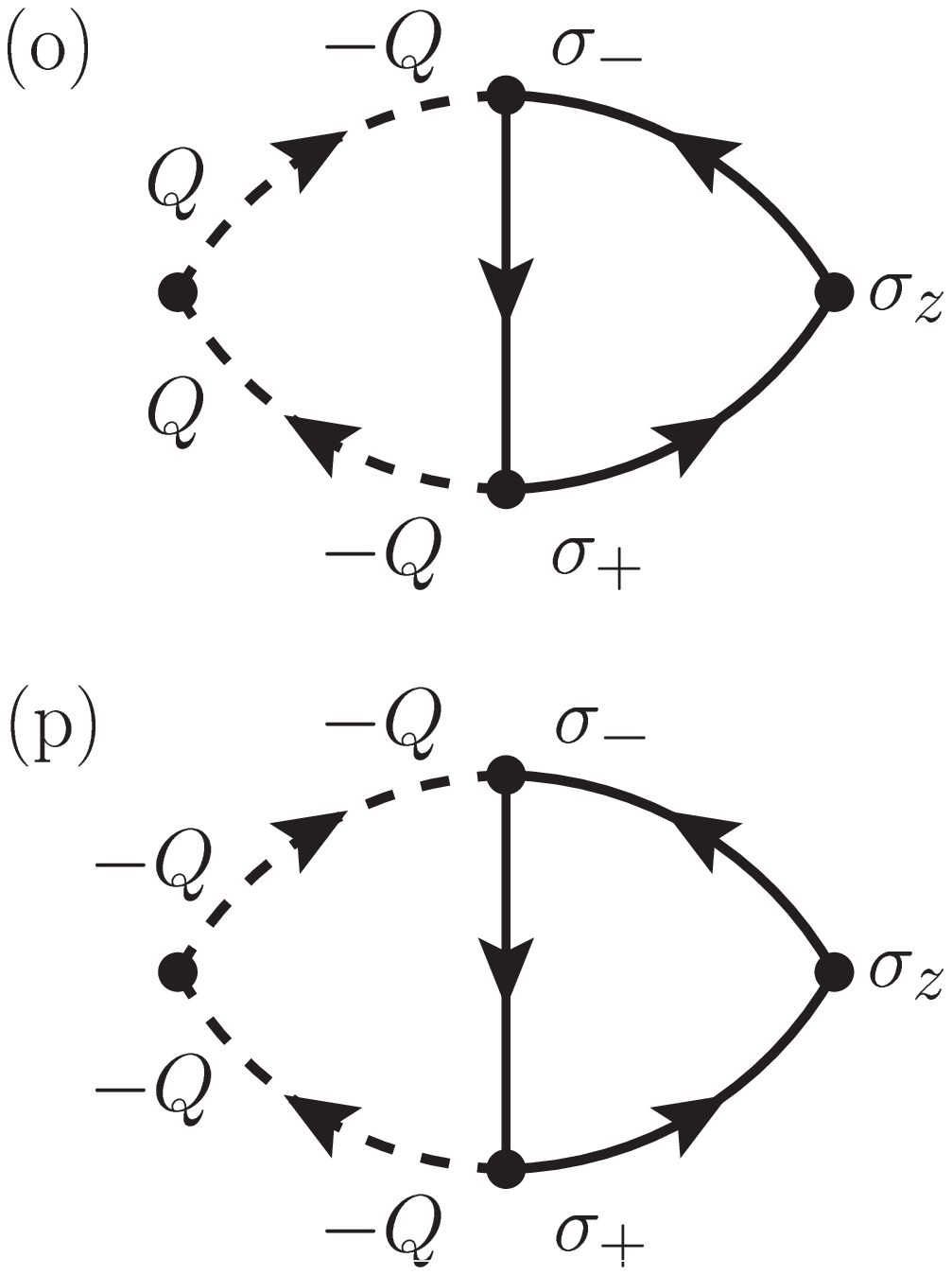} 
%
%
\caption{Feynman diagrams for the phason drag. 
The solid lines and the dashed lines represent electron and phonon Green's functions, respectively, 
and the $\pm Q$ attached to the dashed lines represent the subscripts of phonon Green's functions
${\mathcal O}_{mn}$ and $\tilde{\mathcal O}_{mn}$ with $m, n=\pm$. 
For diagrams (e)-(p), only the Pauli matrices and the sign of $Q$ are shown, while the momenta and 
Matsubara frequencies are the same with the corresponding diagram in (a)-(d).}
\label{f1}
\end{figure}

The Feynman diagrams for the phason drag contributions 
$\Phi_{12}^{\rm ph} (i\omega_\lambda)$ of eq.~(\ref{PhasonDCorr}) are shown in Fig.~\ref{f1}. 
The diagrams (a)-(d) give
\begin{eqnarray}
{\rm (a)}: \quad - g_Q^2\frac{(k_{\rm B}T)^2}{L^2} && \sum_{k,q, n,\nu} \hbar \omega_Q c_Q 
\tilde{\mathcal O}_{++} (q, i\omega_\nu) {\mathcal O}_{++} (q, i\omega_\nu + i\omega_\lambda) \cr
&&\times ev_{\rm F} {\rm Tr} \left[ 
{\mathcal G} (k, i\varepsilon_n) \sigma_+ {\mathcal G}(k-q, i\varepsilon_n -i\omega_\nu) \sigma_-
{\mathcal G}(k, i\varepsilon_n+i\omega_\lambda) \sigma_z  \right] \cr
\quad = - g_Q^2\frac{(k_{\rm B}T)^2}{L^2} && \sum_{k,q, n,\nu} \hbar \omega_Q c_Q 
\tilde{\mathcal O}_{++} (q, i\omega_\nu) {\mathcal O}_{++} (q, i\omega_\nu + i\omega_\lambda)  \cr
&&\times ev_{\rm F} \biggl\{ 
{\mathcal G}_{++} (k, i\varepsilon_n) {\mathcal G}_{--}(k-q, i\varepsilon_n -i\omega_\nu) 
{\mathcal G}_{++} (k, i\varepsilon_n+i\omega_\lambda) \cr
&&- {\mathcal G}_{-+} (k, i\varepsilon_n) {\mathcal G}_{--}(k-q, i\varepsilon_n -i\omega_\nu) 
{\mathcal G}_{+-} (k, i\varepsilon_n+i\omega_\lambda) 
\biggr\}, \cr
{\rm (b)}: \quad  g_Q^2\frac{(k_{\rm B}T)^2}{L^2} && \sum_{k,q, n,\nu} \hbar \omega_Q c_Q 
\tilde{\mathcal O}_{+-} (q, i\omega_\nu) {\mathcal O}_{-+} (q, i\omega_\nu + i\omega_\lambda) \cr
&&\times ev_{\rm F} {\rm Tr} \left[ 
{\mathcal G} (k, i\varepsilon_n) \sigma_+ {\mathcal G}(k-q, i\varepsilon_n -i\omega_\nu) \sigma_-
{\mathcal G}(k, i\varepsilon_n+i\omega_\lambda) \sigma_z \right] \cr
\quad =  g_Q^2\frac{(k_{\rm B}T)^2}{L^2} && \sum_{k,q, n,\nu} \hbar \omega_Q c_Q 
\tilde{\mathcal O}_{+-} (q, i\omega_\nu) {\mathcal O}_{-+} (q, i\omega_\nu + i\omega_\lambda)  \cr
&&\times ev_{\rm F} \biggl\{  
{\mathcal G}_{++} (k, i\varepsilon_n) {\mathcal G}_{--}(k-q, i\varepsilon_n -i\omega_\nu) 
{\mathcal G}_{++} (k, i\varepsilon_n+i\omega_\lambda) \cr
&&- {\mathcal G}_{-+} (k, i\varepsilon_n) {\mathcal G}_{--}(k-q, i\varepsilon_n -i\omega_\nu) 
{\mathcal G}_{+-} (k, i\varepsilon_n+i\omega_\lambda) 
\biggr\},
\nonumber
\end{eqnarray}
\begin{eqnarray}
{\rm (c)}: \quad - g_Q^2\frac{(k_{\rm B}T)^2}{L^2} && \sum_{k,q, n,\nu} \hbar \omega_Q c_Q 
\tilde{\mathcal O}_{++} (q, i\omega_\nu) {\mathcal O}_{++} (q, i\omega_\nu + i\omega_\lambda) \cr
&&\times ev_{\rm F} {\rm Tr} \left[ 
{\mathcal G} (k, i\varepsilon_n) \sigma_- {\mathcal G}(k+q, i\varepsilon_n +i\omega_\nu+i\omega_\lambda)  \sigma_+ 
{\mathcal G} (k, i\varepsilon_n+i\omega_\lambda) \sigma_z\right]
\cr
\quad = - g_Q^2\frac{(k_{\rm B}T)^2}{L^2} && \sum_{k,q, n,\nu} \hbar \omega_Q c_Q 
\tilde{\mathcal O}_{++} (q, i\omega_\nu) {\mathcal O}_{++} (q, i\omega_\nu + i\omega_\lambda)  \cr
&&\times ev_{\rm F} \bigg\{ 
{\mathcal G}_{+-} (k, i\varepsilon_n) {\mathcal G}_{++}(k+q, i\varepsilon_n +i\omega_\nu+i\omega_\lambda) 
{\mathcal G}_{-+} (k, i\varepsilon_n+i\omega_\lambda) \cr
&&- {\mathcal G}_{--} (k, i\varepsilon_n) {\mathcal G}_{++}(k+q, i\varepsilon_n +i\omega_\nu+i\omega_\lambda) 
{\mathcal G}_{--} (k, i\varepsilon_n+i\omega_\lambda)
\biggr\}, \cr
{\rm (d)}: \quad g_Q^2\frac{(k_{\rm B}T)^2}{L^2} && \sum_{k,q, n,\nu} \hbar \omega_Q c_Q 
\tilde{\mathcal O}_{+-} (q, i\omega_\nu) {\mathcal O}_{-+} (q, i\omega_\nu + i\omega_\lambda) \cr
&&\times ev_{\rm F} {\rm Tr} \left[ 
{\mathcal G} (k, i\varepsilon_n) \sigma_- {\mathcal G}(k+q, i\varepsilon_n +i\omega_\nu+i\omega_\lambda)  \sigma_+ 
{\mathcal G} (k, i\varepsilon_n+i\omega_\lambda) \sigma_z \right]
\cr
\quad = g_Q^2\frac{(k_{\rm B}T)^2}{L^2} && \sum_{k,q, n,\nu} \hbar \omega_Q c_Q 
\tilde{\mathcal O}_{+-} (q, i\omega_\nu) {\mathcal O}_{-+} (q, i\omega_\nu + i\omega_\lambda)  \cr
&&\times ev_{\rm F} \bigg\{ 
{\mathcal G}_{+-} (k, i\varepsilon_n) {\mathcal G}_{++}(k+q, i\varepsilon_n +i\omega_\nu+i\omega_\lambda) 
{\mathcal G}_{-+} (k, i\varepsilon_n+i\omega_\lambda) \cr
&&- {\mathcal G}_{--} (k, i\varepsilon_n) {\mathcal G}_{++}(k+q, i\varepsilon_n +i\omega_\nu+i\omega_\lambda) 
{\mathcal G}_{--} (k, i\varepsilon_n+i\omega_\lambda)
\biggr\},
\label{PhiPD1}
\end{eqnarray}
respectively. 
$\omega_{\pm Q+q}$, $c_{\pm Q+q}$, and $g_{\pm Q+q}$
are approximated as $\omega_Q, \pm c_Q$, and $g_Q$, respectively. 
Here it is to be noted that the phonon propagators are \lq\lq directed" as fermions, i.e., ${\mathcal O}_{mn}(q,i\omega_\nu)$ and 
$\tilde {\mathcal O}_{mn}(q,i\omega_\nu)$ are Fourier transform of 
\begin{eqnarray}
{\mathcal O}_{mn}(q,\tau) &=& -\langle T_\tau 
[b^{\phantom{\dagger}}_{mQ+q} (\tau) \left( b^\dagger_{nQ+q} (0) + b^{\phantom{\dagger}}_{-nQ-q} (0)\right) ] \rangle, \cr
\tilde {\mathcal O}_{mn}(q,\tau) &=& -\langle T_\tau 
[\left( b^{\phantom{\dagger}}_{mQ+q} (\tau) + b^{{\dagger}}_{-mQ-q} (\tau) \right) b^\dagger_{nQ+q} (0) ] \rangle,
\end{eqnarray}
respectively, with $m, n=\pm$.  

For diagrams (e)-(p), the $\pm$ signs for $\mathcal G_{mn}$, $\mathcal O_{mn}$, 
and $\tilde {\mathcal O}_{mn}$ are different from those in 
diagrams (a)-(d), while the momenta and Matsubara frequencies are the same. 
Noting that the electronic part is common in the diagrams (e) and (f), for example, we obtain
\begin{eqnarray}
{\rm (e)}+{\rm (f)}: \quad - g_Q^2\frac{(k_{\rm B}T)^2}{L^2} && \sum_{k,q, n,\nu} \hbar \omega_Q c_Q \biggl\{
\tilde{\mathcal O}_{++} (q, i\omega_\nu) {\mathcal O}_{+-} (q, i\omega_\nu + i\omega_\lambda) 
-\tilde {\mathcal O}_{+-} (q, i\omega_\nu) {\mathcal O}_{--} (q, i\omega_\nu + i\omega_\lambda) \biggr\} \cr
&&\times ev_{\rm F} \biggl\{  
{\mathcal G}_{++} (k, i\varepsilon_n) {\mathcal G}_{-+}(k-q, i\varepsilon_n -i\omega_\nu) 
{\mathcal G}_{-+} (k, i\varepsilon_n+i\omega_\lambda) \cr
&&- {\mathcal G}_{-+} (k, i\varepsilon_n) {\mathcal G}_{-+}(k-q, i\varepsilon_n -i\omega_\nu) 
{\mathcal G}_{--} (k, i\varepsilon_n+i\omega_\lambda) 
\biggr\}, \cr
{\rm (g)}+{\rm (h)}: \quad - g_Q^2 \frac{(k_{\rm B}T)^2}{L^2} && \sum_{k,q, n,\nu} \hbar \omega_Q c_Q \biggl\{
\tilde{\mathcal O}_{++} (q, i\omega_\nu) {\mathcal O}_{+-} (q, i\omega_\nu + i\omega_\lambda) 
-\tilde {\mathcal O}_{+-} (q, i\omega_\nu) {\mathcal O}_{--} (q, i\omega_\nu + i\omega_\lambda) \biggr\} \cr
&&\times ev_{\rm F} \bigg\{ 
{\mathcal G}_{++} (k, i\varepsilon_n) {\mathcal G}_{-+}(k+q, i\varepsilon_n +i\omega_\nu+i\omega_\lambda) 
{\mathcal G}_{-+} (k, i\varepsilon_n+i\omega_\lambda) \cr
&&- {\mathcal G}_{-+} (k, i\varepsilon_n) {\mathcal G}_{-+}(k+q, i\varepsilon_n +i\omega_\nu+i\omega_\lambda) 
{\mathcal G}_{--} (k, i\varepsilon_n+i\omega_\lambda)
\biggr\}, 
\label{PhiPD2}
\end{eqnarray}
\begin{eqnarray}
{\rm (i)}+{\rm (j)}: \quad - g_Q^2\frac{(k_{\rm B}T)^2}{L^2} && \sum_{k,q, n,\nu} \hbar \omega_Q c_Q \biggl\{
\tilde{\mathcal O}_{-+} (q, i\omega_\nu) {\mathcal O}_{++} (q, i\omega_\nu + i\omega_\lambda) 
-\tilde {\mathcal O}_{--} (q, i\omega_\nu) {\mathcal O}_{-+} (q, i\omega_\nu + i\omega_\lambda) \biggr\} \cr
&&\times ev_{\rm F} \biggl\{ 
{\mathcal G}_{+-} (k, i\varepsilon_n) {\mathcal G}_{+-}(k-q, i\varepsilon_n -i\omega_\nu) 
{\mathcal G}_{++} (k, i\varepsilon_n+i\omega_\lambda) \cr
&&- {\mathcal G}_{--} (k, i\varepsilon_n) {\mathcal G}_{+-}(k-q, i\varepsilon_n -i\omega_\nu) 
{\mathcal G}_{+-} (k, i\varepsilon_n+i\omega_\lambda) 
\biggr\}, \cr
{\rm (k)}+{\rm (l)}: \quad - g_Q^2 \frac{(k_{\rm B}T)^2}{L^2} && \sum_{k,q, n,\nu} \hbar \omega_Q c_Q \biggl\{
\tilde{\mathcal O}_{-+} (q, i\omega_\nu) {\mathcal O}_{++} (q, i\omega_\nu + i\omega_\lambda) 
-\tilde {\mathcal O}_{--} (q, i\omega_\nu) {\mathcal O}_{-+} (q, i\omega_\nu + i\omega_\lambda) \biggr\} \cr
&&\times ev_{\rm F} \bigg\{  
{\mathcal G}_{+-} (k, i\varepsilon_n) {\mathcal G}_{+-}(k+q, i\varepsilon_n +i\omega_\nu+i\omega_\lambda) 
{\mathcal G}_{++} (k, i\varepsilon_n+i\omega_\lambda) \cr
&&- {\mathcal G}_{--} (k, i\varepsilon_n) {\mathcal G}_{+-}(k+q, i\varepsilon_n +i\omega_\nu+i\omega_\lambda) 
{\mathcal G}_{+-} (k, i\varepsilon_n+i\omega_\lambda)
\biggr\}, 
\label{PhiPD3}
\end{eqnarray}
\begin{eqnarray}
{\rm (m)}+{\rm (n)}: \quad - g_Q^2\frac{(k_{\rm B}T)^2}{L^2} && \sum_{k,q, n,\nu} \hbar \omega_Q c_Q \biggl\{
\tilde{\mathcal O}_{-+} (q, i\omega_\nu) {\mathcal O}_{+-} (q, i\omega_\nu + i\omega_\lambda) 
-\tilde {\mathcal O}_{--} (q, i\omega_\nu) {\mathcal O}_{--} (q, i\omega_\nu + i\omega_\lambda) \biggr\} \cr
&&\times ev_{\rm F} \biggl\{  
{\mathcal G}_{+-} (k, i\varepsilon_n) {\mathcal G}_{++}(k-q, i\varepsilon_n -i\omega_\nu) 
{\mathcal G}_{-+} (k, i\varepsilon_n+i\omega_\lambda) \cr
&&- {\mathcal G}_{--} (k, i\varepsilon_n) {\mathcal G}_{++}(k-q, i\varepsilon_n -i\omega_\nu) 
{\mathcal G}_{--} (k, i\varepsilon_n+i\omega_\lambda) 
\biggr\}, \cr
{\rm (o)}+{\rm (p)}: \quad - g_Q^2 \frac{(k_{\rm B}T)^2}{L^2} && \sum_{k,q, n,\nu} \hbar \omega_Q c_Q \biggl\{
\tilde{\mathcal O}_{-+} (q, i\omega_\nu) {\mathcal O}_{+-} (q, i\omega_\nu + i\omega_\lambda) 
-\tilde {\mathcal O}_{--} (q, i\omega_\nu) {\mathcal O}_{--} (q, i\omega_\nu + i\omega_\lambda) \biggr\} \cr
&&\times ev_{\rm F} \bigg\{ 
{\mathcal G}_{++} (k, i\varepsilon_n) {\mathcal G}_{--}(k+q, i\varepsilon_n +i\omega_\nu+i\omega_\lambda) 
{\mathcal G}_{++} (k, i\varepsilon_n+i\omega_\lambda) \cr
&&- {\mathcal G}_{-+} (k, i\varepsilon_n) {\mathcal G}_{--}(k+q, i\varepsilon_n +i\omega_\nu+i\omega_\lambda) 
{\mathcal G}_{+-} (k, i\varepsilon_n+i\omega_\lambda)
\biggr\}.
\label{PhiPD4}
\end{eqnarray}

The \lq\lq directed" phonon propagators, $\mathcal O_{mn}$ and $\tilde {\mathcal O}_{mn}$ 
are obtained from the Dyson equations shown in Appendix B. 
Substituting the explicit form of ${\mathcal G}_{mn}$ in eq.~(\ref{GreenF}) 
and using the relationship between ${\mathcal O}_{mn}, \tilde{\mathcal O}_{mn}$ and ${\mathcal D}_{mn}$ 
obtained in eq.~(\ref{SimpleRel}), the total of eq.~(\ref{PhiPD1}) and eqs.~(\ref{PhiPD2})-(\ref{PhiPD4}) becomes
\begin{eqnarray}
\Phi_{12}^{\rm ph} (i\omega_\lambda) &=& -e \hbar v_{\rm F}  \omega_Q c_Q g_Q^2 \frac{(k_{\rm B}T)^2}{L^2} \sum_{k,q, n,\nu} 
\frac{i\omega_\nu + \hbar\omega_Q}{2\hbar \omega_Q} 
\frac{i\omega_\nu +i\omega_\lambda + \hbar\omega_Q}{2\hbar \omega_Q} 
\frac{2}{(i \varepsilon_n)^2-E_k^2} \frac{1}{(i\varepsilon_n + i\omega_\lambda)^2-E_k^2} \cr
&&\times \biggl[ 
\biggl\{ {\mathcal D}_{++} (q, i\omega_\nu) {\mathcal D}_{++} (q, i\omega_\nu + i\omega_\lambda) 
- e^{-4i\phi} {\mathcal D}_{+-} (q, i\omega_\nu) {\mathcal D}_{+-} (q, i\omega_\nu + i\omega_\lambda) \biggr\} \cr
&&\times \biggl\{ ( f(-q, -i\omega_\nu) - f(q, i\omega_\nu+i\omega_\lambda) ) 
\left\{i \varepsilon_n (i \varepsilon_n+i\omega_\lambda)+ \xi_k^2 - \Delta_0^2\right\} \cr
&&- (g(-q, -i\omega_\nu) - g(q, i\omega_\nu+i\omega_\lambda) ) 
\xi_k (2i \varepsilon_n+i\omega_\lambda) 
\biggr\} \cr
&&+ \biggl\{ {\mathcal D}_{++} (q, i\omega_\nu) {\mathcal D}_{+-} (q, i\omega_\nu + i\omega_\lambda) 
-{\mathcal D}_{+-} (q, i\omega_\nu) {\mathcal D}_{++} (q, i\omega_\nu + i\omega_\lambda) \biggr\} \cr
&&\times \biggl\{ -\frac{i\omega_\lambda (\Delta^*)^2}{(i\varepsilon_n-i\omega_\nu)^2 -E_{k-q}^2}
-\frac{i\omega_\lambda (\Delta^*)^2}{(i\varepsilon_n+i\omega_\nu+i\omega_\lambda )^2-E_{k+q}^2}\biggr\} \biggr], 
\end{eqnarray}
where $f(q, i\omega_\nu)$ and $g(q, i\omega_\nu)$ are defined in eq.~(\ref{fgfunctDef}).
The last terms with $i \omega_\lambda (\Delta^*)^2$ can be neglected in the following 
since it is proportional to $(i\omega_\lambda)^2$. 
Finally, using the phason and amplitude propagators defined in eq.~(\ref{DefPhasonAmp}), we obtain eq.~(\ref{Phixx}).

\section{Straightforward derivation of Eq.~(\ref{CxqLowest})}

Putting $q=0$ and $i\omega_\lambda=0$ in the electron Green's functions in eq.~(\ref{PhiPD1}), 
the electronic part of $\Phi_{12}^{\rm ph}(i\omega_\lambda)$ corresponding to the diagrams 
in Fig.~\ref{f1}(a) and (c) becomes
\begin{eqnarray}
\sum_k {\rm Tr} \biggl[ 
&&{\mathcal G} (k, i\varepsilon_n) \sigma_+ {\mathcal G}(k, i\varepsilon_n -x) \sigma_-
{\mathcal G}(k, i\varepsilon_n) \sigma_z  \cr
+&&
{\mathcal G} (k, i\varepsilon_n) \sigma_- {\mathcal G}(k, i\varepsilon_n +x) \sigma_+
{\mathcal G}(k, i\varepsilon_n) \sigma_z \biggr], 
\label{FukuTech1}
\end{eqnarray}
where $i\omega_\nu$ is replaced by $x$.
To evaluate the trace in eq.~(\ref{FukuTech1}), we use
\begin{eqnarray}
\sigma_+ \left( \begin{array}{cc} a & b \cr c & d \end{array} \right) \sigma_- &=& 
\left( \begin{array}{cc} d & 0 \cr 0 & 0 \end{array} \right) = \frac{d}{2} (\sigma_0+\sigma_z) , \cr
\sigma_- \left( \begin{array}{cc} a & b \cr c & d \end{array} \right) \sigma_+ &=& 
\left( \begin{array}{cc} 0 & 0 \cr 0 & a \end{array} \right) = \frac{a}{2} (\sigma_0-\sigma_z),
\end{eqnarray}
where $\sigma_0$ is the $2\times 2$ unit matrix.
When $x=0$, the trace in eq.~(\ref{FukuTech1}) becomes
\begin{eqnarray}
&&\frac{1}{2} {\rm Tr} \biggl[ 
{\mathcal G} (k, i\varepsilon_n) \frac{i\varepsilon_n - \xi_k}{(i\varepsilon_n)^2-E_k^2}  (\sigma_0+\sigma_z) 
{\mathcal G}(k, i\varepsilon_n) \sigma_z  \cr
&&+
{\mathcal G} (k, i\varepsilon_n)  \frac{i\varepsilon_n + \xi_k}{(i\varepsilon_n)^2-E_k^2}  (\sigma_0-\sigma_z) 
{\mathcal G}(k, i\varepsilon_n) \sigma_z \biggr] \cr
&=&\frac{1}{(i\varepsilon_n)^2-E_k^2} {\rm Tr} \biggl[
{\mathcal G} (k, i\varepsilon_n) (i\varepsilon_n \sigma_0 - \xi_k \sigma_z) 
{\mathcal G}(k, i\varepsilon_n) \sigma_z  \biggr] \cr
&=&\frac{2}{[(i\varepsilon_n)^2-E_k^2]^3} \biggl[ i\varepsilon_n \times (2i\varepsilon_n \xi_k) 
- \xi_k \times \left( (i\varepsilon_n)^2 + \xi_k^2 - \Delta_0^2 \right) \biggr].
\end{eqnarray}
This vanishes since the last expression is odd with respect to $k$. 
The lowest order with respect to $x$ becomes, in a similar way, 
\begin{eqnarray}
&x& {\rm Tr} \biggl[ 
{\mathcal G} (k, i\varepsilon_n) \sigma_+ {\mathcal G}^2(k, i\varepsilon_n) \sigma_-
{\mathcal G}(k, i\varepsilon_n) \sigma_z  
- 
{\mathcal G} (k, i\varepsilon_n) \sigma_- {\mathcal G}^2(k, i\varepsilon_n) \sigma_+
{\mathcal G}(k, i\varepsilon_n) \sigma_z \biggr], \cr
&=&\frac{x}{[(i\varepsilon_n)^2-E_k^2]^2} {\rm Tr} \biggl[
{\mathcal G} (k, i\varepsilon_n) \left\{ - 2i\varepsilon_n \xi_k \sigma_0
+ \left( (i\varepsilon_n)^2+\xi_k^2 + \Delta_0^2 \right) \sigma_z \right\} 
{\mathcal G}(k, i\varepsilon_n) \sigma_z  \biggr] \cr
&=&\frac{2x}{[(i\varepsilon_n)^2-E_k^2]^4} \biggl[ - 2i\varepsilon_n \xi_k  \times (2i\varepsilon_n \xi_k) 
+ \left( (i\varepsilon_n)^2+\xi_k^2 + \Delta_0^2 \right) \times \left( (i\varepsilon_n)^2 + \xi_k^2 - \Delta_0^2 \right) \biggr] \cr
&=&\frac{2x}{[(i\varepsilon_n)^2-E_k^2]^4} \biggl[ 
\left( (i\varepsilon_n)^2-\xi_k^2 \right)^2 - \Delta_0^4 \biggr],
\end{eqnarray}
which leads to Eq.~(\ref{CxqLowest}).
Other contributions in Fig.~\ref{f1} are treated similarly. 

\section{Calculation of $D(T)$}

\begin{eqnarray}
D(T) &=& - \frac{k_{\rm B}T}{L} \sum_{k,n} 
\frac{(i\varepsilon_n)^2 - \xi_k^2 + \Delta_0^2  }{[(i\varepsilon_n)^2-E_k^2]^3} \cr
&=& \frac{1}{L} \sum_k 
\oint \frac{dz}{2\pi i} f(z) 
\left[ \frac{1}{(z^2-E_k^2)^2} +\frac{2\Delta_0^2}{(z^2-E_k^2)^3} \right] \cr
&=& - \frac{1}{L} \sum_k 
\sum_{\pm} \left[ \pm \frac{\Delta_0^2 f''(\pm E_k)}{8 E_k^3}  
+ \frac{2E_k^2-3\Delta_0^2}{8 E_k^4} f'(\pm E_k) \mp \frac{2E_k^2-3\Delta_0^2}{8 E_k^5} f(\pm E_k) \right] ,
\label{DT1}
\end{eqnarray}
where $f(\varepsilon) = 1/(e^{\beta\varepsilon}+1)$, i.e., Fermi distribution function with $\mu=0$.
At $T=0$, we have
\begin{eqnarray}
D(0) &= - \frac{1}{L} \sum_k 
\frac{2E_k^2-3\Delta_0^2}{8 E_k^5}.
\label{DT2}
\end{eqnarray}
Using a relation 
\begin{equation}
\frac{d}{dk} \left( \frac{k}{E_k^3} \right)=-\frac{2E_k^2-3\Delta_0^2}{E_k^5}, 
\end{equation}
eq.~(\ref{DT2}) becomes
\begin{equation}
D(0) = \frac{{k_{\rm F}}}{8\pi E_{k_{\rm F}}^3}, 
\end{equation}
where we have used the range of $k$ is $|k|<k_{\rm F}$ as assumed in eq.~(\ref{MFHamilton}). 

\section{Phase Hamiltonian and phason propagator}

The phason propagator eq.~(\ref{PhasonandAmp2}) at $T=0$ is derived from the former studies 
based on the phase Hamiltonian approach. 
The model of phason coupled to randomly distributed impurities is given by\cite{Fukuyama76,FukuLee}
\begin{equation}
H_0 = \pi \hbar v' \int dx \left[ p^2 + \frac{1}{4\pi^2} \left( \frac{v}{v'} \right)^2 (\nabla \phi)^2 \right],
\label{AppendixFno1}
\end{equation}
and 
\begin{equation}
H' = {V_0 \rho_0} \sum_i \cos[ Q R_i + \phi(R_i)],
\label{PhaseHIMP}
\end{equation}
where $v$ is the phason velocity given in eq.~(\ref{PhasonVelocityXX}), 
and $v'=v^2/|v_{\rm F}|$. 
The first one is field theory for phase variable $\phi(x)$ while the second represent the coupling to impurities of CDW 
expressed in terms of $\phi(x)$\ogata{, which is derived from the impurity Hamiltonian, 
$H'$, in eq.~(\ref{FrohlichHamil}) and the charge density, $\rho(x)$, in eq.~(\ref{RhoCDWdef}), 
assuming that $v(r-R_i)=V_0 \delta(x-R_i)$. } 
Electrical conductivity for uniform electric field with finite frequency $\omega$, $\sigma(\omega)$, is 
given as follows by noting that the current density operator is represented as $-(e/\pi) \partial \phi(x,t)/\partial t$.\cite{LRA,Fukuyama76}
\begin{equation}
\sigma(\omega) = - \frac{i\omega}{2}\left( \frac{e}{\pi} \right)^2 L {\mathcal D}(0, 0; i\omega_n) 
\biggl|_{i\omega_n \rightarrow \hbar\omega+i\delta},
\label{FukuOld1}
\end{equation}
where phason Green's function is defined by 
\begin{equation}
{\mathcal D}(q, q'; i\omega_n) 
= 2\int_{0}^\beta d\tau e^{i\omega_n \tau} \langle T_\tau [\phi_q(\tau) \phi_{-q'}(0)] \rangle,
\label{FukuOld2}
\end{equation}
and $\phi_q$ is the Fourier transform of $\phi(x)$ defined as
\begin{equation}
\phi_q = \frac{1}{L} \int dx e^{-iqx} \phi(x). 
\end{equation}
It is to be noted that the formulation in \cite{Fukuyama76} and \cite{FukuLee}, in particular the convention of $\omega$, 
is changed in accordance with the present framework. 


Here we note that 
the phonon operator $b^{\phantom{\dagger}}_{Q+q}(\tau) + b^{{\dagger}}_{-Q-q} (\tau)$ can be expressed 
in terms of the phase and amplitude variables as 
\begin{equation}
b^{\phantom{\dagger}}_{Q+q} (\tau) + b^{{\dagger}}_{-Q-q} (\tau)
\sim \frac{\sqrt{L}}{g_Q} \left( \Delta_0+\delta \Delta_q (\tau) \right) e^{i(\phi + \delta\phi_q(\tau))},
\end{equation}
where, $\delta\Delta_q(\tau)$ represents the modulation of the amplitude and $\delta\phi_q(\tau)$ represents the 
deviation from the constant $\phi$.  
Substituting this expression into the phonon propagator in eq.~(\ref{PGreenDef}), and 
using the expansion $e^{i\delta\phi_q(\tau)}\sim 1+i\delta\phi_q(\tau)$, 
we obtain 
\begin{eqnarray}
{\mathcal D}_{++}(q,\tau) 
&&\sim - \frac{L}{g_Q^2} \langle T_\tau [\delta \Delta_q (\tau) \delta \Delta_{-q}(0) ] \rangle 
- \frac{L\Delta_0^2}{g_Q^2} \langle T_\tau  [\delta\phi_q(\tau) \delta\phi_{-q}(0)] \rangle, \cr
{\mathcal D}_{+-}(q,\tau) 
&&\sim - \frac{L}{g_Q^2} e^{2i\phi} \langle T_\tau [\delta \Delta_q (\tau) \delta \Delta_{-q}(0) ] \rangle 
+ \frac{L\Delta_0^2}{g_Q^2} e^{2i\phi} \langle T_\tau  [\delta\phi_q(\tau) \delta\phi_{-q}(0)] \rangle.
\end{eqnarray}
Hence we obtain by noting $\phi_q = \phi+\delta\phi_q$, 
\begin{equation}
P(q, i\omega_n) = {\mathcal D}_{++}(q, i\omega_n)-e^{-2i\phi}
{\mathcal D}_{+-}(q, i\omega_n) 
= - 2 \frac{L\Delta_0^2}{g_Q^2} \int_{0}^\beta d\tau e^{i\omega_n \tau} \langle T_\tau  [\delta\phi_q(\tau) \delta\phi_{-q}(0)] \rangle
= - \frac{L\Delta_0^2}{g_Q^2} {\mathcal D}(q, i\omega_n), \label{F_9}
\end{equation}
which proves the equivalence between eq.~(\ref{phasonSigma}) and eq.~(\ref{FukuOld1})

In a clean system without disorder this phason propagator governed by (\ref{AppendixFno1}) is given by 
${\mathcal D}(q, q'; i\omega_n) = \delta_{q,q'} {\mathcal D}_0(q,i\omega_n)$ with
\begin{equation}
{\mathcal D}_0 (q, i\omega_n) = \frac{1}{L} \ \frac{4\pi \hbar v'}{\omega_n^2+ \hbar^2 v^2 q^2}.
\end{equation}
By noting $v= \left( {m}/{m^*} \right)^{1/2} |v_{\rm F}|$ with $m^*$ being the effective mass of phason mode, 
$\sigma(\omega)$, eq.~(\ref{FukuOld1}), in this case for spatially uniform electric field ($q=0$) is 
\begin{equation}
\sigma(\omega) = \frac{i n_e e^2}{m^* \omega}, \label{SuperCondXX}
\end{equation}
which is the same as LRA. 
Equation (\ref{SuperCondXX}) is considered to be the manifestation of Fr\"ohlich 
superconductivity in the Peierls phase without disorder. 

\begin{figure}
\includegraphics[width=10cm]{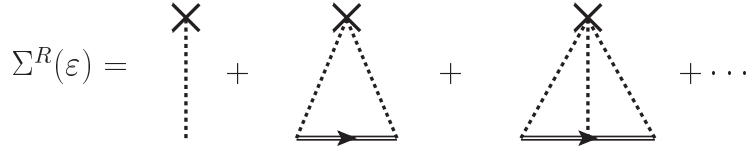} 
\caption{Feynman diagrams in $t$-matrix approximation. 
Crosses represent the impurity potential and the double solid lines are the phason propagator. 
}
\label{TMatrix}
\end{figure}

Effects of impurity scattering to the phason propagator are given by the self-energy correction, $\Gamma$, defined by 
\begin{equation}
\langle {\mathcal D}(q, q'; i\omega_n) \rangle_{\rm av} = \delta_{q+q'} [{\mathcal D}_0(q)^{-1} -\Gamma]^{-1} 
= \delta_{q+q'} {\mathcal D}(q, i\omega_n).
\end{equation}
The $t$-matrix approximation to $\Gamma$ is given by the processes in Fig.~\ref{TMatrix}. 
As clarified in Ref.~\cite{FukuLee} effects of impurity pinning can be classified typically into weak and strong, 
characterized by the parameter $\varepsilon= V_0 \rho_0/n_i \hbar |v_{\rm F}|$.  
We focus to the case of weak pinning $\varepsilon<<1$ for generality. 
In this case, the first and second order terms in Fig.~\ref{TMatrix} are sufficient. 
The first order contribution is given by
\begin{equation}
\Gamma_1 = \frac{V_0 \rho_0}{2L} \sum_i \cos[ Q R_i + \phi(R_i)]. 
\end{equation}
This contribution is vanishing if the phase is rigid, i.e., spatially constant. 
However there is gain of energy due to spatial distortions of $\phi$ reflecting distribution of impurities 
leading to domains with characteristic size $L_0$ which is given by (similar to random walk problem) 
\begin{equation}
\Gamma_1 = -\frac{1}{2} V_0 \rho_0  (n_i L_0)^{1/2} /L_0. \label{TMatGamm1}
\end{equation}
Here the size $L_0$ is to be determined by optimizing the energy gain (\ref{TMatGamm1}) against the energy loss due to 
spatial distortion of the phase represented by the second term of (\ref{AppendixFno1}), 
leading to $(n_iL_0)^{-1}=(\alpha\pi \varepsilon)^{2/3}$
where the parameter $\alpha$ reflects the way of phase distortion of the order of $\pi$. 
The study in \cite{FukuLee} has indicated that $\alpha= 3^3/2^5$ is the best choice. 
This is the essence of impurity pinning. The second order contribution is given by 
\begin{eqnarray}
\Gamma_2 &=& \left( \frac{V_0 \rho_0}{2L} \right)^2 \sum_{q\ne 0} {\mathcal D}(q, \hbar\omega)  
\sum_{i,j} e^{iq(R_i-R_j)} \cos[ Q R_i + \phi(R_i)]\cos[ Q R_j + \phi(R_j)] \cr
&=& \frac{n_i}{2} \left( \frac{V_0 \rho_0}{2} \right)^2 2\pi \frac{v'}{v} \left( -\hbar^2 \omega^2 -4\pi \hbar v'\Gamma \right)^{-1/2}.
\label{TMatGamm2}
\end{eqnarray}

The self-consistent equation for $\Gamma=\Gamma_1+\Gamma_2$ given by (\ref{TMatGamm1}) and (\ref{TMatGamm2}) lead to 
\begin{equation}
G= -2\alpha^{1/3} + (-y^2 -G)^{-1/2}, 
\end{equation}
where $G=4\pi \hbar v'\Gamma/\gamma^2$, and $y=\hbar \omega/\gamma$ 
with $\gamma=(\pi \varepsilon)^{2/3} \omega_0$ and $\omega_0=n_i \hbar v$. 
For low frequency, $y<1$, the solution of the self-consistent equation for $G$ with proper choice of parameter charactering 
effects of impurity scattering for causality to be satisfied is found to be $G \sim -a_0 + ia_1 y$ with 
$a_0=1/2^{2/3} = 0.630$  and $a_1=(2^{4/3}/3)^{1/2} = 0.916$,
which leads to 
\begin{equation}
{\mathcal D} (q, \hbar \omega+i\delta) = \frac{1}{L} \ \frac{4\pi \hbar v'}{-(\hbar\omega+i\delta)^2+ \hbar^2 v^2 q^2 + g_0 - i\hbar \omega g_1}.
\label{TmatFinalX}
%
\end{equation}
where $g_0=\gamma^2 a_0$ and $g_1 = \gamma a_1$. 
Eqs. (\ref{TmatFinalX}), (\ref{F_9}), together with $v'=v^2/|v_{\rm F}|$, $v= (X/(1+X))^{1/2} |v_{\rm F}|$, and
$X = \omega_Q g_Q^2 / 2\pi |v_{\rm F}| \Delta_0^2$ 
lead to (\ref{PhasonandAmp2}).

\def\journal#1#2#3#4{#1 {\bf #2}, #3 (#4)}
\def\PR{Phys.\ Rev.}
\def\PRB{Phys.\ Rev.\ B}
\def\PRL{Phys.\ Rev.\ Lett.}
\def\JPSJ{J.\ Phys.\ Soc.\ Japan}
\def\PTP{Prog.\ Theor.\ Phys.}
\def\JPCS{J.\ Phys.\ Chem.\ Solids}


\begin{thebibliography}{9}

\bibitem{ExpRev2} For a review, see M.\ D.\ Mahan, {\it Good Thermoelectrics},
\journal{Solid State Physics}{51}{81}{1997}.

\bibitem{ExpRev} K.\ Behnia \lq\lq Fundamentals of Thermoelectricity'' 
(Oxford University Press 2015).

\bibitem{Kubo} R.\ Kubo, 
{\it Statistical-Mechanical Theory of Irreversible Processes. I. 
General Theory and Simple Applications to Magnetic and Conduction Problems},
\journal{\JPSJ}{12}{570}{1957}. 

\bibitem{Luttinger} J.\ M.\ Luttinger, 
{\it Theory of Thermal Transport Coefficients}, 
\journal{\PR}{135}{A1505}{1964}. 

\bibitem{OF1} M.\ Ogata and H.\ Fukuyama, 
{\it Theory of Spin Seebeck Effects in a Quantum Wire}, 
\journal{\JPSJ}{86}{094703}{2017}. 

\bibitem{OFSB} M.\ Ogata and H.\ Fukuyama, 
{\it Range of Validity of Sommerfeld-Bethe Relation Associated with Seebeck Coefficient and Phonon Drag Contribution}, 
\journal{\JPSJ}{88}{074703}{2019}. 

\bibitem{Matsuura} H.\ Matsuura, H.\ Maebashi, M.\ Ogata, and H.\ Fukuyama, 
{\it Effect of Phonon Drag on Seebeck Coefficient Based on Linear Response Theory: Application to FeSb$_2$}, 
\journal{\JPSJ}{88}{074601}{2019}. 

\bibitem{YamaFuku1} T.\ Yamamoto and H.\ Fukuyama, 
{\it Possible High Thermoelectric Power in Semiconducting Carbon Nanotubes --A Case Study of Doped One-Dimensional Semiconductors--},
\journal{\JPSJ}{87}{024707}{2018}. 

\bibitem{YamaFuku2} T.\ Yamamoto and H.\ Fukuyama, 
{\it Bipolar Thermoelectric Effects in Semiconducting Carbon Nanotubes: Description in Terms of One-Dimensional Dirac Electrons}, 
\journal{\JPSJ}{87}{114710}{2018}. 

\bibitem{Gurevich} L.\ Gurevich, 
{\it Thermoelectric properties of conductors. I}, \journal{J.\ Phys.\ }{6}{477}{1945}.

\bibitem{Herring} C.\ Herring, 
{\it Theory of the Thermoelectric Power of Semiconductors}, 
\journal{Phys.\ Rev. }{96}{1163}{1954}.

\bibitem{Mahan} G.\ D.\ Mahan, L.\ Lindsay, and D.\ A.\ Broido, 
{\it The Seebeck coefficient and phonon drag in silicon},
\journal{J.\ Appl.\  Phys.\ }{116}{245102}{2014}.

\bibitem{Zhou} J.\ Zhou, B.\ Liao, B.\ Qiu, S.\ Huberman, K.\ Esfarjani, M.\ S.\ Dresselhaus, and G.\ Chen, 
{\it Ab initio optimization of phonon drag effect for lower-temperature thermoelectric energy conversion}, 
\journal{Proc.\ Natl Acad.\ Sci.\ }{112}{14777}{2015}.

\bibitem{LRA} P.\ A.\ Lee, T.\ M.\ Rice, and P.\ W.\ Anderson, 
{\it Conductivity from charge or spin density waves}, 
\journal{Solid State Commun.}{14}{703}{1974}.

\bibitem{Coleman} L.\ B.\ Coleman, M.\ J.\ Cohen, D.\ J.\ Sandman, F.\ G.\ Yamagishi, A.\ F.\ Garito, 
and A.\ J.\ Heeger, 
{\it Superconducting fluctuations and the Peierls instability in an organic solid}, 
\journal{Solid State Commun.}{12}{1125}{1973}.

\bibitem{Cohen} M.\ J.\ Cohen, L.\ B.\ Coleman, A.\ F.\ Garito, and A.\ J.\ Heeger, 
{\it Electrical conductivity of tetrathiofulvalinium tetracyanoquinodimethan (TTF) (TCNQ)}, 
\journal{\PRB}{10}{1298}{1974}.

\bibitem{Kurihara} H.\ Yoshimoto and S.\ Kurihara, 
{\it Thermal Transport Properties of a Charge Density Wave}, 
\journal{\JPSJ}{75}{014601}{2006}. 

\bibitem{Fukuyama76} H.\ Fukuyama, 
{\it Pinning in Peierls-Fr\" ohlich State and Conductivity}, 
\journal{\JPSJ}{41}{513}{1976}. 

\bibitem{FukuLee} H.\ Fukuyama and P.\ A.\ Lee,
{\it Dynamics of the charge-density wave. I. Impurity pinning in a single chain}, \journal{\PRB}{17}{535}{1978}. 

\bibitem{TakayamaFuku} H.\ Fukuyama and H.\ Takayama, 
{\it Electronic Properties of Inorganic Quasi-One-Dimensional Materials}, ed.\ by P.\ Monceau
(1985, D. Reidel Publishing Company).

\bibitem{Baumann} K.\ Baumann, 
{\it Quantum theory of transport coefficients. II}, 
\journal{Annals of Phys.}{23}{221}{1963}. 

\bibitem{Sakuma} Similar processes have been considered in the following studies on the magnon drag phenomena.
D.\ Miura, and A.\ Sakuma, 
{\it Microscopic Theory of Magnon-Drag Thermoelectric Transport in Ferromagnetic Metals}, 
\journal{\JPSJ}{81}{113602}{2012}.

\bibitem{Okuma} N.\ Okuma and K.\ Nomura, 
{\it Microscopic derivation of magnon spin current in a topological insulator/ferromagnet heterostructure}, 
\journal{\PRB}{95}{115403}{2017}.

\bibitem{Kohno} T.\ Yamaguchi and H.\ Kohno, 
{\it Microscopic Theory of Spin-Wave Spin Torques Induced by Temperature Gradient}, \journal{\JPSJ}{86}{063706}{2017}.

\bibitem{Kohno2} Y.\ Imai and H.\ Kohno, 
{\it Theory of Cross-correlated Electron-Magnon Transport Phenomena: Case of Magnetic Topological Insulator}, 
\journal{\JPSJ}{87}{073709}{2018}.


\bibitem{Baggioli} e.g., M.\ Baggioli and A.\ Zaccone, 
{\it Universal Origin of Boson Peak Vibrational Anomalies in Ordered Crystals and in Amorphous Materials}, 
\journal{\PRL}{122}{145501}{2019}.


\bibitem{AndersonLeeSaitoh} P.\ W.\ Anderson, P.\ A.\ Lee, and M.\ Saitoh, 
{\it Remarks on giant conductivity in TTF-TCNQ}, 
\journal{Solid State Commun.}{13}{595}{1973}.


\bibitem{Fukuyama78} H.\ Fukuyama, 
{\it Commensurability Pinning versus Impurity Pinning of One-Dimensional Charge Density Wave}, 
\journal{\JPSJ}{45}{1474}{1978}. 

\bibitem{energyLand} H.\ Fukuyama, J.\ Kishine, and M.\ Ogata, 
{\it Energy Landscape of Charge Excitations in the Boundary Region between Dimer-Mott and Charge Ordered States in Molecular Solids}, 
\journal{\JPSJ}{86}{123706}{2017}. 


\bibitem{Kwak} J.\ F.\ Kwak, P.\ M.\ Chaikin, A.\ A.\ Russel, A.\ F.\ Garito, and A.\ J.\ Heeger,
{\it Anisotropic thermoelectric power of TTF-TCNQ}, 
\journal{Solid State Commun.}{16}{729}{1975}.

\end{thebibliography}
\end{document}